\documentclass[letterpaper,11pt]{article}
\pdfoutput=1

\usepackage[T1]{fontenc}
\usepackage[utf8]{inputenc}

\usepackage{graphicx}
\usepackage{epstopdf}
\usepackage{multicol}
\usepackage{lmodern}

\usepackage{jheppub}
\usepackage{microtype}   
\usepackage{multirow, amsthm}
\usepackage{braket}
\usepackage{mathrsfs}
\usepackage{tikz}
\usepackage[caption=false]{subfig}
\usepackage{xspace}
\usepackage{xcolor}
\usepackage{float,color}
\usepackage[separate-uncertainty=true]{siunitx}
\usepackage[ruled,norelsize]{algorithm2e}
\usepackage[countmax]{subfloat}
\usepackage{accents}
\usepackage{physics}
\usepackage[capitalise]{cleveref}    

\newcommand{\fastjet}[1]{\textsc{FastJet\xspace #1}}
\newcommand{\pythia}[1]{\textsc{Pythia\xspace #1}}
\newcommand{\mg}[1]{\textsc{MadGraph5\_aMC@NLO\xspace #1}}

\newlength{\dhatheight}

\providecommand{\href}[2]{#2}
\usepackage{comment}

\definecolor{darkred}{rgb}{0.5,0.0,0.0}
\definecolor{darkblue}{rgb}{0.0,0.0,0.9}
\definecolor{darkerblue}{rgb}{0.0,0.0,0.5}
\definecolor{darkgreen}{rgb}{0.0,0.5,0.0}
\definecolor{black}{rgb}{0.0,0.0,0.0}
\definecolor{brown}{rgb}{0.6,0.4,0.2}

\DeclareRobustCommand{\Ref}[1]{Ref.~\cite{#1}}
\DeclareRobustCommand{\Refs}[1]{Refs.~\cite{#1}}

\providecommand{\ord}{O}

\DeclareSIUnit{\nb}{\nano\barn}
\DeclareSIUnit{\pb}{\pico\barn}
\DeclareSIUnit{\fb}{\femto\barn}
\DeclareSIUnit{\year}{yr}

\newcommand{\pt}{\ensuremath{p_T}\xspace}

\title{\boldmath Boosting $H\to b\bar b$ with Machine Learning
}
\author[1,2]{Joshua Lin,}
\author[3]{Marat Freytsis,}
\author[4,5]{Ian Moult,}
\author[1]{and Benjamin Nachman}
\affiliation[1]{\normalsize Physics Division, Lawrence Berkeley National Laboratory, Berkeley, CA 94720, USA}
\affiliation[2]{\normalsize Department of Physics, University of California, Berkeley, Berkeley, CA 94720, USA}
\affiliation[3]{\normalsize Institute of Theoretical Science, University of Oregon, Eugene, OR 97403, USA}
\affiliation[4]{Berkeley Center for Theoretical Physics, University of California, Berkeley, CA 94720, USA}
\affiliation[5]{Theoretical Physics Group, Lawrence Berkeley National Laboratory, Berkeley, CA 94720, USA}

\emailAdd{joshua.z.lin@berkeley.edu}
\emailAdd{freytsis@uoregon.edu}
\emailAdd{ianmoult@lbl.gov}
\emailAdd{bpnachman@lbl.gov}

\abstract{
High-\pt Higgs production at hadron colliders provides a direct probe of the internal structure of the $gg \to H$ loop with the $H \to b\bar{b}$ decay offering the most statistics due to the large branching ratio.  Despite the overwhelming QCD background, recent advances in jet substructure have put the observation of the $gg \to H \to b\bar{b}$ channel at the LHC within the realm of possibility.  In order to enhance the sensitivity to this process, we develop a two-stream convolutional neural network, with one stream acting on jet information and one using global event properties.  The neural network significantly increases the discovery potential of a Higgs signal, both for high-\pt Standard Model production as well for possible beyond the Standard Model contributions.  Unlike most studies for boosted hadronically decaying massive particles, the boosted Higgs search is unique because double $b$-tagging rejects nearly all background processes that do not have two hard prongs.  In this context --- which goes beyond state-of-the-art two-prong tagging --- the network is studied to identify the origin of the additional information leading to the increased significance.  The procedures described here are also applicable to related final states where they can be used to identify additional sources of discrimination power that are not being exploited by current techniques.
}

\begin{document} 
\maketitle
\flushbottom

\section{Introduction}
\label{sec:intro}

Even though it has been over five years since the discovery of the Higgs boson~\cite{Aad:2012tfa,Chatrchyan:2012xdj}, the final state with the largest branching ratio ($\mathcal{B}(H \to b\bar{b}) \approx \SI{58}{\percent}$~\cite{Heinemeyer:2013tqa}) has not been probed with great precision.  This process is difficult to measure due to large and nearly irreducible background processes --- only recently has $(V)H\to b\bar{b}$ been confirmed~\cite{ATLAS-CONF-2018-036,Aaboud:2017xsd,Sirunyan:2017elk}.  However, many interesting and largely untested features of high-\pt Higgs production~\cite{Grojean:2013nya,Azatov:2013xha,Buschmann:2014twa,Schlaffer:2014osa,Azatov:2016xik} are challenging to probe with cleaner final states such as $H \to \gamma\gamma$ or $H \to ZZ^* \to 4\ell$ due to their low branching ratios.\footnote{Studies of the recently observed $t\bar{t}H$ production~\cite{Aaboud:2017rss,Sirunyan:2018mvw} provide complementary information on the underlying physics of $pp \to H$ production.}   Boosted $H \to b\bar{b}$ provides access to the highest \pt Higgs bosons at the Large Hadron Collider (LHC); if they can be measured with good precision, a door leading beyond the Standard Model (BSM) could be opened.

Major advances in the use of jet substructure and machine learning techniques have revolutionized the ability to look for hadronic signals in increasingly extreme regions of phase space.\footnote{For a review of recent theoretical and experimental progress, see~\Refs{Larkoski:2017jix,Asquith:2018igt}.}  Most analyses that exploit the hadronic decays of boosted heavy particles have so far used modern tools to purify the event selection but not to directly identify the main objects of interest.  However, pioneering work by the ATLAS and CMS collaborations have used these techniques to directly measure boosted particle cross-sections~\cite{Aad:2014bla,Aad:2014haa,Aaboud:2018eqg,Sirunyan:2017dgc} and to search directly for BSM particles~\cite{Sirunyan:2017dnz,Sirunyan:2017nvi,Aaboud:2018zba}.  In particular, the CMS collaboration has used single jets to search for the boosted $H \to b\bar{b}$ decay~\cite{Sirunyan:2017dgc}, the first experimental result on the subject since the idea was originally proposed in \Ref{Butterworth:2008iy} (albeit exclusively in the $VH$ channel), as well as combining this analysis with other differential low-\pt data~\cite{CMS-PAS-HIG-17-028}.  One reason for the long delay between conception and practical results was the development of advanced techniques for grooming~\cite{Dasgupta:2013ihk,Dasgupta:2013via,Larkoski:2014wba}, $2$-prong tagging~\cite{Thaler:2010tr,Moult:2016cvt,Larkoski:2013eya,Larkoski:2014gra}, jet four-vector calibrations~\cite{ATLAS-CONF-2017-063,CMS-DP-2017-026}, and boosted $b$-tagging~\cite{ATLAS-CONF-2012-100,CMS:2013vea,ATL-PHYS-PUB-2014-014,ATL-PHYS-PUB-2015-035,CMS-PAS-BTV-15-001,Sirunyan:2017ezt,ATLAS-CONF-2016-001,ATLAS-CONF-2016-002}.

The presence of multiple nearby boosted $b$ quarks sets boosted Higgs identification apart from other boosted massive particle classification.  This is because requiring two $b$-tagged subjets inside a larger jet necessarily requires that the parent jet has a two-prong structure.  For boosted massive object identification, most of the jet substructure community has focused on $n$-prong taggers~\cite{Thaler:2010tr,Thaler:2011gf,Larkoski:2013eya,Larkoski:2014gra,Larkoski:2014zma,Larkoski:2015kga,Salam:2016yht,Moult:2016cvt,Komiske:2017aww,Larkoski:2017iuy,Larkoski:2017cqq,Datta:2017lxt}, which are not optimized for cases where $n$-prongs are already present.  By probing the full radiation pattern inside boosted boson decays, \Ref{deOliveira:2015xxd} showed that there is information beyond traditional $n$-prong tagging and even beyond traditional color flow observables~\cite{Gallicchio:2010sw}.  This was also explored in the context of boosted Higgs boson decays in \Ref{Datta:2017lxt}, which identified simple observables that capture the additional information.  However, neither these studies nor the more recent \Ref{Lim:2018toa}, which considered generic quark and gluon jets as a background to $H \to b\bar{b}$, were explicitly predicated on subjet tagging as a baseline and did not probe global information beyond jet substructure.  Until the present work, the full potential of information beyond $n$-prong taggers has not been demonstrated for concrete observables such as cross-sections or BSM coupling limits.\footnote{The latest CMS $b\bar{b}$ and $c\bar{c}$ tagging techniques use machine learning approaches with a large number of particle- and vertex-level inputs~\cite{CMS-DP-2018-046}.  These approaches could learn information beyond $n$-prong tagging.  However, the background source for training is generic quark and gluon jets, not $g \to b\bar{b}$.  The working point used in the boosted $H$ search operates at the \SI{1}{\percent} mis-tag rate, while the rate of $g \to b\bar{b}$ is comparable to or lower than this value (see e.g., \Ref{Abe:1999vw}) so most of the tagger's effort must go to reducing the large non-$g \to b\bar{b}$ background.  Performance studies specifically with $g \to b\bar{b}$ as the background show that the tagger reduces the $g \to b\bar{b}$ background $3\times$ more than the signal~\cite{CMS-PAS-BTV-15-002}.  The equivalent performance shown later in this paper (\cref{fig:SIC_beta}) corresponds to about $16\times$ more $g \to b\bar{b}$ than signal.  These numbers are not directly comparable because the latter is also after mass and two-prong tagging requirements (and is thus conservative).  Therefore, the techniques presented in this paper are using more information, but further studies are required to understand how much more and what type of information is being used.}

Modern machine learning (ML) tools have shown great promise for using low-level~\cite{Cogan:2014oua,Almeida:2015jua,deOliveira:2015xxd,Baldi:2016fql,Barnard:2016qma,Komiske:2016rsd,deOliveira:2017pjk,Kasieczka:2017nvn,Louppe:2017ipp,Pearkes:2017hku,Datta:2017rhs,Komiske:2017ubm,Butter:2017cot,Metodiev:2017vrx,Datta:2017lxt,Haake:2017dpr,Cheng:2017rdo,Egan:2017ojy,Komiske:2017aww,Komiske:2018oaa,Macaluso:2018tck,Chien:2018dfn,Fraser:2018ieu,Andreassen:2018apy,Collins:2018epr,Choi:2018dag,Guo:2018hbv,Lim:2018toa,Monk:2018zsb,Guest:2016iqz,Bhimji:2017qvb,ATL-PHYS-PUB-2017-003,ATL-PHYS-PUB-2017-013,ATL-PHYS-PUB-2017-017,CMS-DP-2017-027,Sirunyan:2017ezt} and high-level~\cite{ATL-PHYS-PUB-2017-004,ATLAS-CONF-2017-064,Duarte:2018ite,Lenz:2017lqo} information to classify hadronic final states at the LHC.  These techniques must be adapted to cope with significant sparsity, large dynamic ranges, multi-channel inputs and data that has no unique representation.  Similar techniques have been demonstrated for full event classification wth low-level~\cite{Bhimji:2017qvb,Louppe:2017ipp} and high-level inputs~\cite{Baldi:2014kfa,Baldi:2014pta,Abdughani:2018wrw}.  In addition to the challenges related to the structure of the data, one of the key challenges for applying state-of-the-art techniques in practice is the need for a background estimation method. As in \Ref{Collins:2018epr}, boosted $H \to b\bar{b}$ has a natural background estimation technique by using the localization of the Higgs boson in the jet mass distribution.  For this reason, the algorithms presented here may already be useful to enhance existing analysis efforts.

In this paper we use deep neural networks to examine the potential of using all the available information in boosted Higgs events. We use a two stream convolutional neural network to combine jet substructure information with global event information, finding significant gains coming from both components demonstrating that the search can be greatly improved. Furthermore, we are able to identify the dominant source of jet substructure discrimination in terms of a simple observable.

This paper is organized as follows.  \Cref{sec:sim} details the ML setup, including the preprocessing and architecture of the neural network. The neural network is then applied to the SM search for boosted $H \to b\bar{b}$ in \cref{sec:higgs_SM}.  Physics beyond the SM could introduce \pt-dependent effects that are enhanced for boosted Higgs bosons.  Implications for the NN classifier on BSM physics are described in \cref{sec:BSM}.  The paper concludes in \cref{sec:conclusions}.

\section{Machine Learning Architecture}
\label{sec:sim}

This section describes our machine learning setup, with a focus on the neural network architecture and preprocessing.

\subsection{Neural Network Architecture}

Our neural network architecture is driven both by physics goals as well as the desire to extract the maximal amount of information from the event. For the boosted $H \to b\bar{b}$ topology, there are two physically distinct components to the events: the substructure of the hardest jet and the global event structure.   Due to the color singlet scalar nature of the Higgs, the radiation pattern within and around the $b\bar{b}$ jet is expected to differ from $g \to b\bar{b}$ jets.  Different production mechanisms can also result in different numbers and orientations of jets in the events.  All of these aspects are investigated.  

\begin{figure}
  \begin{center}
    \includegraphics[width=\textwidth]{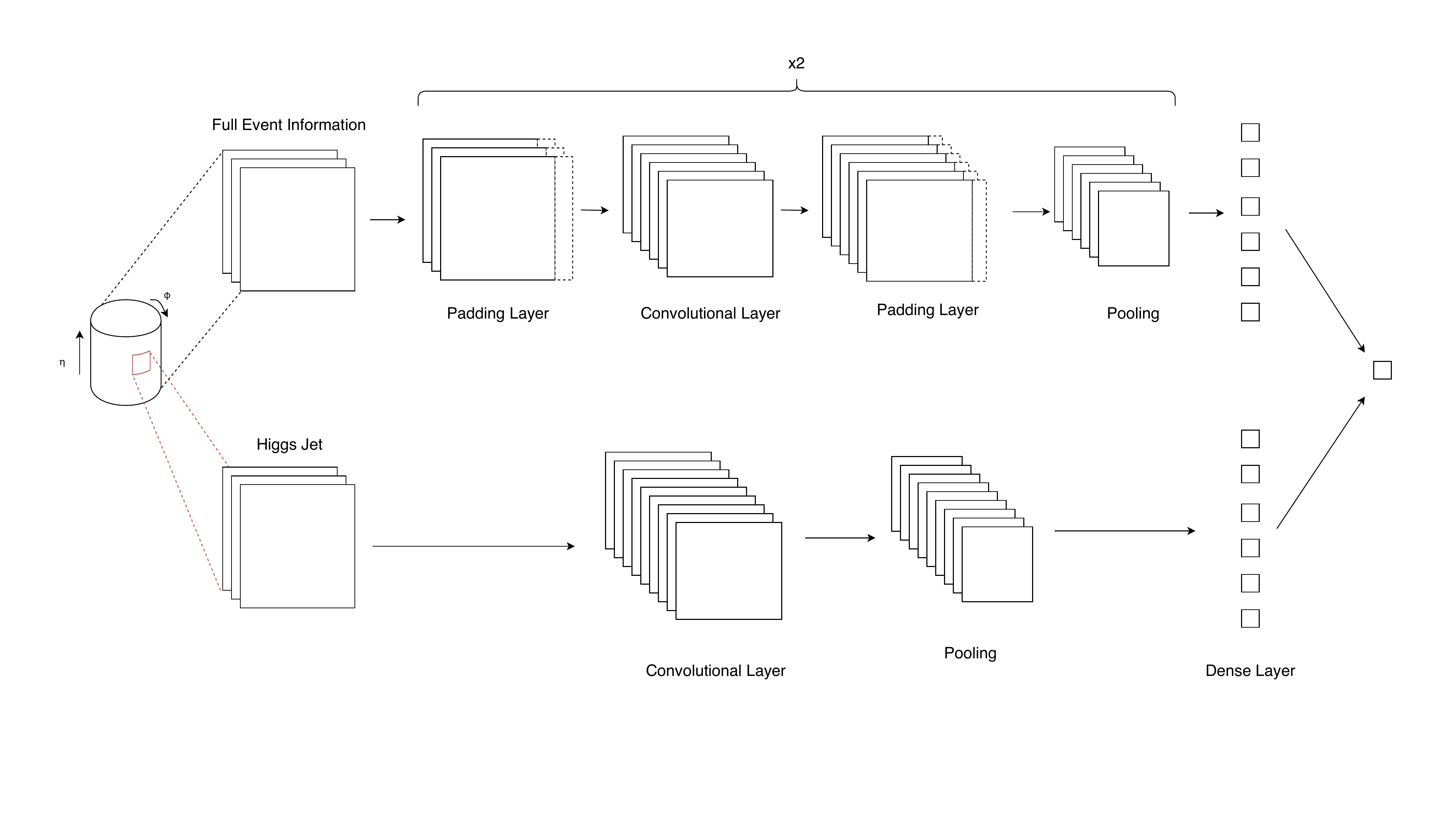}
  \end{center}
  \caption{A schematic of the two-stream CNN used in this study. The first stream uses the full event information, while the second stream uses the jet substructure information. More details on the architecture are provided in the text.}
  \label{fig:NN_architecture}
\end{figure}

To incorporate both local and global information, a two-stream neural network is constructed.  One stream acts on the full event information and the other acts on the image of the Higgs candidate jet. The two streams are then combined.  This setup can be used to separately assess how much discrimination power can be obtained from the substructure and the global event separately, as well as in combination. A schematic of this two-stream architecture is shown in \cref{fig:NN_architecture}.

In order to account for the compact nature of the detector in the $\phi$ direction, we use padding layers that take the leftmost few columns and append them to the right before each convolution (for all convolutional layers), effectively performing convolutions over the cylinder rather than over a square.  Further details related to the image (pre)processing are discussed in \cref{sec:pre_process}.

The details of the convolution and pooling layers of each stream are as follows. Each convolutional filter is $5 \times 5$, and the pooling layers are $2 \times 2$, with rectified linear unit (ReLU) activations, and stride length of 1. The first convolutional layer in each stream has 32 filters, and the second convolutional layer in each stream has 64 filters. The dense layer at the end of each stream has 300 neurons each. Finally, the two dense layers from each stream are fully connected to an output layer of one neuron with sigmoid activation. In total this gives 2.6 million trainable parameters in the network. We used the AdaDelta optimizer \cite{DBLP:journals/corr/abs-1212-5701}, with binary cross entropy as our loss function, and used the relatively simple Early Stopping method as a regularization technique, stopping when the significance improvement of the Higgs measurement at $\pt^\text{min} = \SI{450}{\GeV}$ stopped improving (with a patience of 2 epochs). We arrived at this final model after testing the performance (measured by the significance improvement of the Higgs measurement at $\pt^\text{min} = \SI{450}{\GeV}$) using different optimizers (AdaDelta \cite{DBLP:journals/corr/abs-1212-5701}, AdaGrad \cite{Duchi:2011:ASM:1953048.2021068}, Adam \cite{DBLP:journals/corr/KingmaB14}), different activation functions (mainly testing ReLU against leaky ReLU), and regularization (dropout \cite{JMLR:v15:srivastava14a} vs.\ Early Stopping). Our training was performed using the \textsc{Keras}~\cite{keras} Python neural network library with \textsc{Tensorflow}~\cite{tensorflow} backend, on Nvidia GeForce 1080 Ti GPUs.

\subsection{Inputs and Preprocessing}
\label{sec:pre_process}

The inputs to our neural network are jet images~\cite{Cogan:2014oua}. For each event, an image is created for each stream: one image is the full event image and the other is the image of the hardest jet (that has been double $b$-tagged). Both images are $40 \times 40$ pixels. For the jet image, the range (in $\eta$--$\phi$ space) is $2R \times 2R$ where $R = 0.8$ is the radius of the jet. The full event image covers effectively the entire $\eta$--$\phi$ cylinder ($|\eta| < 5$). Inspired by \Ref{Komiske:2016rsd}, both the jet and event image have three channels analogous to the RGB channels of a color image.  The pixel intensity for the three channels correspond to the sum of the charged particle \pt, the sum of the neutral particle \pt, and the number of charged particles.  As the neutral particle \pt is particularly sensitive to pileup, additional studies without this channel are included in the results.

To ensure that the neural network is not learning spacetime symmetries, and to reduce the size of the input streams, the jet images are preprocessed in a similar way to previous studies, see e.g., \Refs{deOliveira:2015xxd,Komiske:2016rsd}.  In particular, all of the images are normalized (sum of intensities is unity) and standardized (zero-centered and divided by standard deviations).  Prior to these steps, the jet images are also rotated so that the two subjets are aligned along the same axis in every image~\cite{deOliveira:2015xxd,Cogan:2014oua}.  Details about the subjet identification and $b$-tagging are discussed in \cref{sec:higgs_SM_standard}.

\section{Boosting Standard Model Higgs Tagging}
\label{sec:higgs_SM}

This section studies the neural network performance in the context of improving the significance for the Standard Model boosted $H \to b\bar{b}$ search.

\subsection{Simulation Setup and Validation}
\label{sec:higgs_SM_standard}

Simulated $pp$ collisions at $\sqrt{s} = \SI{13}{\TeV}$ are generated using \mg{2.6.2}~\cite{Alwall:2014hca} for the hard processes and showered with \pythia{8.226}~\cite{Sjostrand:2007gs}.\footnote{In our study, we only use events showered by \pythia{8.226}~\cite{Sjostrand:2007gs}. Although it is true that by switching to a different showering program such as Herwig \cite{Bahr:2008pv} or Sherpa \cite{Gleisberg:2008ta} the quantitative improvement shown in Sec.~\ref{sec:higgs_SM_ML} may change, the qualitative gains from the Machine Learning should be robust and motivates a complete analysis to be undertaken by the LHC collaborations. Previous studies with deep learning on images have observed that raw classifier performance varies when different generators are used for testing but give nearly the same result when the testing models are the same (even if the training ones are different). \cite{Komiske:2016rsd}} Background events are generated using two, three and four jet events (\texttt{pp > jj}, \texttt{pp > jjj} and \texttt{pp > jjjj}) matched using the MLM approach~\cite{Mangano:2006rw}.  In order to include finite top mass effects (and BSM contributions in \cref{sec:BSM}) signal events are generated at one-loop order (\texttt{pp > Hj [QCD]} and \texttt{pp > Hjj [QCD]}), which in this case corresponds to the leading contribution.  The overlap between the real emission from the matrix element and the parton shower is also accounted for using the MLM algorithm.  Higher order amplitudes for the signal process with full mass dependence are now becoming available~\cite{Chen:2016zka,Neumann:2016dny,Bonciani:2016qxi,Jones:2018hbb,Lindert:2018iug,Kudashkin:2017skd,Neumann:2018bsx}.  These updates could slightly modify the numerical results, but should not change the conclusions and would primarily effect the overall rate and not the features exploited by the machine learning approach, which are primarily associated with the radiation pattern in the jet and the global event. Furthermore, in these studies, higher loop finite top mass effects are found to be flat at high \pt and therefore do not significantly modify the shape of the \pt spectrum.

\begin{figure}
  \begin{center}
    \includegraphics[width=8.5cm]{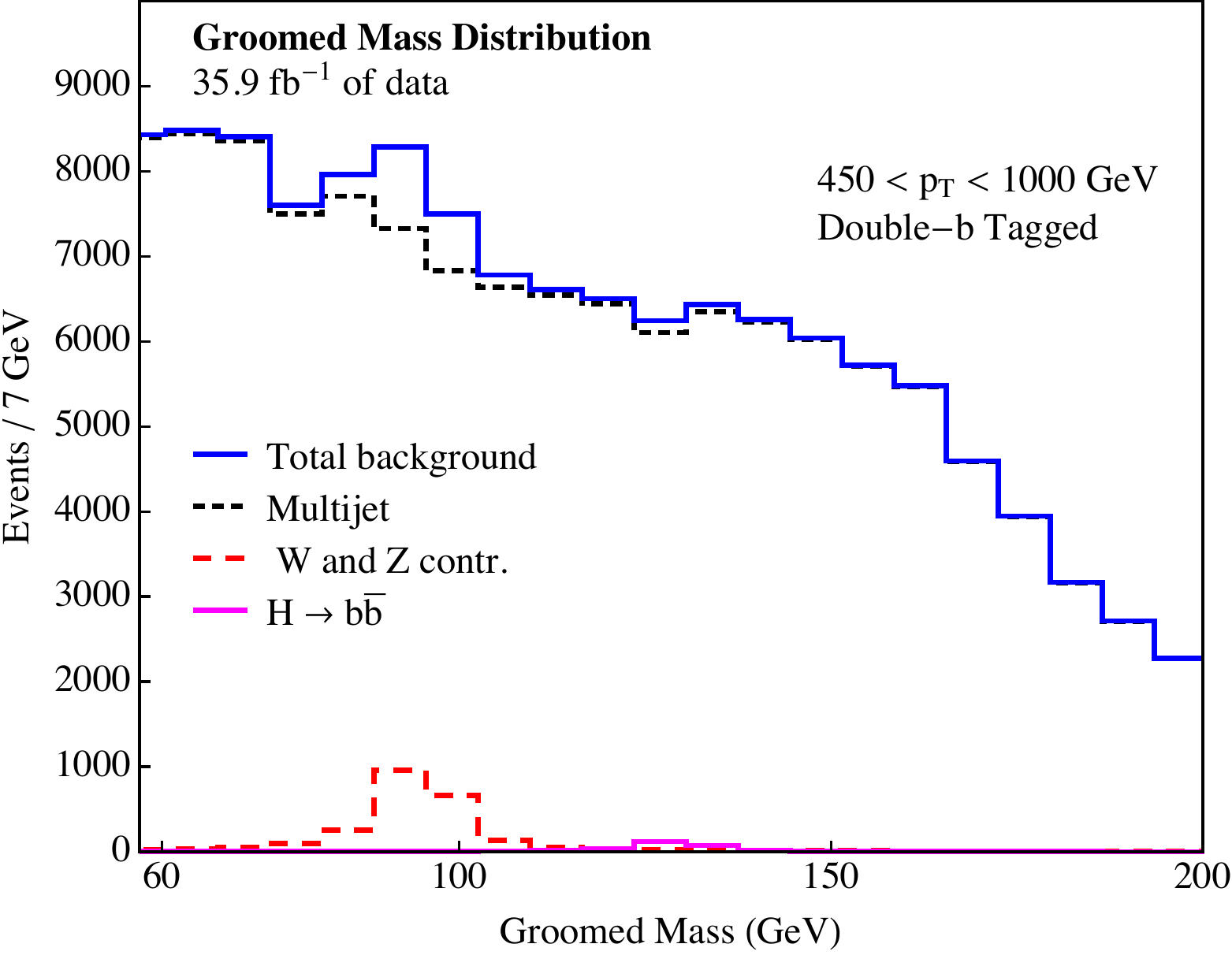}
  \end{center}
  \caption{Groomed mass distribution for signal and background (including leading order vector boson contributions). The small Higgs boson signal is visible near $m_{\text{SD}} \sim \SI{125}{\GeV}$. Note that the simulation statistical uncertainty is comparable to the data statistical uncertainty.  While not ideal, this does not qualitatively change the conclusions.}
  \label{fig:cms_verify}
\end{figure}

Events are clustered and analyzed using \fastjet{3.2.1}~\cite{Cacciari:2011ma} and the \fastjet{contrib} extensions. Following a CMS-like analysis~\cite{Sirunyan:2017dgc}, jets are clustered with $R = 0.8$ anti-$k_t$ jets~\cite{Cacciari:2008gp}, which are groomed with the soft drop algorithm~\cite{Larkoski:2014wba} with $\beta=0$ and $z_\text{cut}=0.1$.\footnote{With this choice of $\beta$, the algorithm is the same as the precursor modified mass drop tagger~\cite{Dasgupta:2013ihk} approach.  It is called soft drop (SD) in this paper in order to avoid confusion, as this is the name used in the CMS analysis.} Candidate Higgs jets are required to have transverse momentum $\pt > \SI{450}{\GeV}$, and satisfy a double $b$-tag.  In general, $b$-tagging performance depends heavily on the exact experimental implementation and is detector specific.  In this analysis, we use an approach similar to subjet $b$-tagging in ATLAS\footnote{The CMS approach does not explicitly reconstruct subjets.  See \cref{sec:conclusions} for additional discussion on this point.}~\cite{Aaboud:2018xwy} which can be mimicked at particle-level while assuming \SI{100}{\percent} $b$-tagging efficiency and infinitely good rejection.  This introduces an $\ord(1)$ correction to the cross-section, but does not qualitatively change the conclusions.  The subjets of the large-$R$ jets are ghost-associated~\cite{Cacciari:2008gn} $R=0.2$ anti-$k_t$ jets.  Such jets are declared $b$-tagged if they have a ghost-associated $B$ hadron with $\pt > \SI{5}{\GeV}$.  In addition to $b$-tagging, the leading double $b$-tagged jet (the Higgs candidate) is required to have $-6.0 < \rho < -2.1$ ($\rho = \log(m_{\text{SD}}^2/\pt^2)$).  This is chosen following \Ref{Sirunyan:2017dgc} to avoid the deeply nonperturbative region as well as finite cone limitations in the jet clustering, although no re-optimization of this range was performed. Finally, the two-prong observable $N_2$ \cite{Moult:2016cvt} is required to be $\leq 0.4$.  There is little dependence on the exact $N_2$ requirement, likely in part because of the two ($b$-tagged) subjet requirement. \Cref{fig:cms_verify} shows the $m_\text{SD}$ distribution after applying the above selections.  The overall rate, relative rates between processes, and general trends agree with the CMS analysis in \Ref{Sirunyan:2017dgc}.

Since the goal of this paper is to emphasize the possible gains for this search using ML, we have made a number of simplifying assumptions, and therefore the exact reproduction of the CMS analysis is not our primary concern. We believe that none of these assumptions significantly change our quantitative conclusions, but they should be revisited with the full analyses in ATLAS and CMS.  In particular, the $t\bar{t}$ background is ignored, the background fit is simplified, experimental effects relating to track reconstruction and $b$-tagging are ignored, and as mentioned above, the Higgs cross-section is only computed at NLO.  The top background is small but comparable to the Higgs signal, and since we have consistently ignored it for the pseudo-data and background, any residual contribution is a subleading effect from modeling uncertainty.  Tracks reconstructed by ATLAS and CMS are excellent proxies for charged particles, though there are percent-level differences resulting from material interactions and pattern recognition ambiguities.  These effects, as well as pileup, can slightly degrade $b$-tagging performance~\cite{Sirunyan:2017ezt,ATL-PHYS-PUB-2017-013}.  Once again, this is important to account for when setting a precise limit, but would not change the relative gains presented here.

\subsection{Machine Learning Results}
\label{sec:higgs_SM_ML}

Having validated the setup specified in \cref{sec:higgs_SM_standard} against the public CMS results~\cite{Sirunyan:2017dgc}, the simulated events are now used as input to our two stream convolutional neural network to identify whether additional discrimination power can be obtained from the jet substructure, jet superstructure, and other global event properties.

Network training proceeds with \num{50000} signal and background events passing the selection criteria from \cref{sec:higgs_SM_standard}. The training-validation-test split that we used was \SI{50}{\percent}--\SI{25}{\percent}--\SI{25}{\percent}.  There is no requirement on the jet mass, as the entire spectrum is used to evaluate the significance.  In practice, this could make traditional data-driven background estimation techniques more complex to use, though there have also been many techniques proposed to preserve the mass distribution~\cite{Dolen:2016kst,Moult:2017okx,Aguilar-Saavedra:2017rzt,Shimmin:2017mfk}.

The neural network performance is quantified using the significance improvement characteristic (SIC) curve.  Such a curve is approximately equal to $\epsilon/\sqrt{\epsilon_b}$ and quantifies the gain in significance over the baseline selection.  Following a CMS-like analysis~\cite{Sirunyan:2017dgc}, the full significance is calculated using a binned likelihood fit treating the bin counts as Poisson-distributed random variables.  This procedure assumes that the results are dominated by statistical uncertainties, which will always be true for the highest \pt bins.  Data statistical uncertainties account for over half of the total uncertainty in \Ref{Sirunyan:2017dgc}, so this is a valid approximation.  There is no fit to determine the background shape, which is taken directly from the simulation.  Once again, this is valid in the statistics limited regime.  

\begin{figure}
  \begin{center}
    \includegraphics[width=8.5cm]{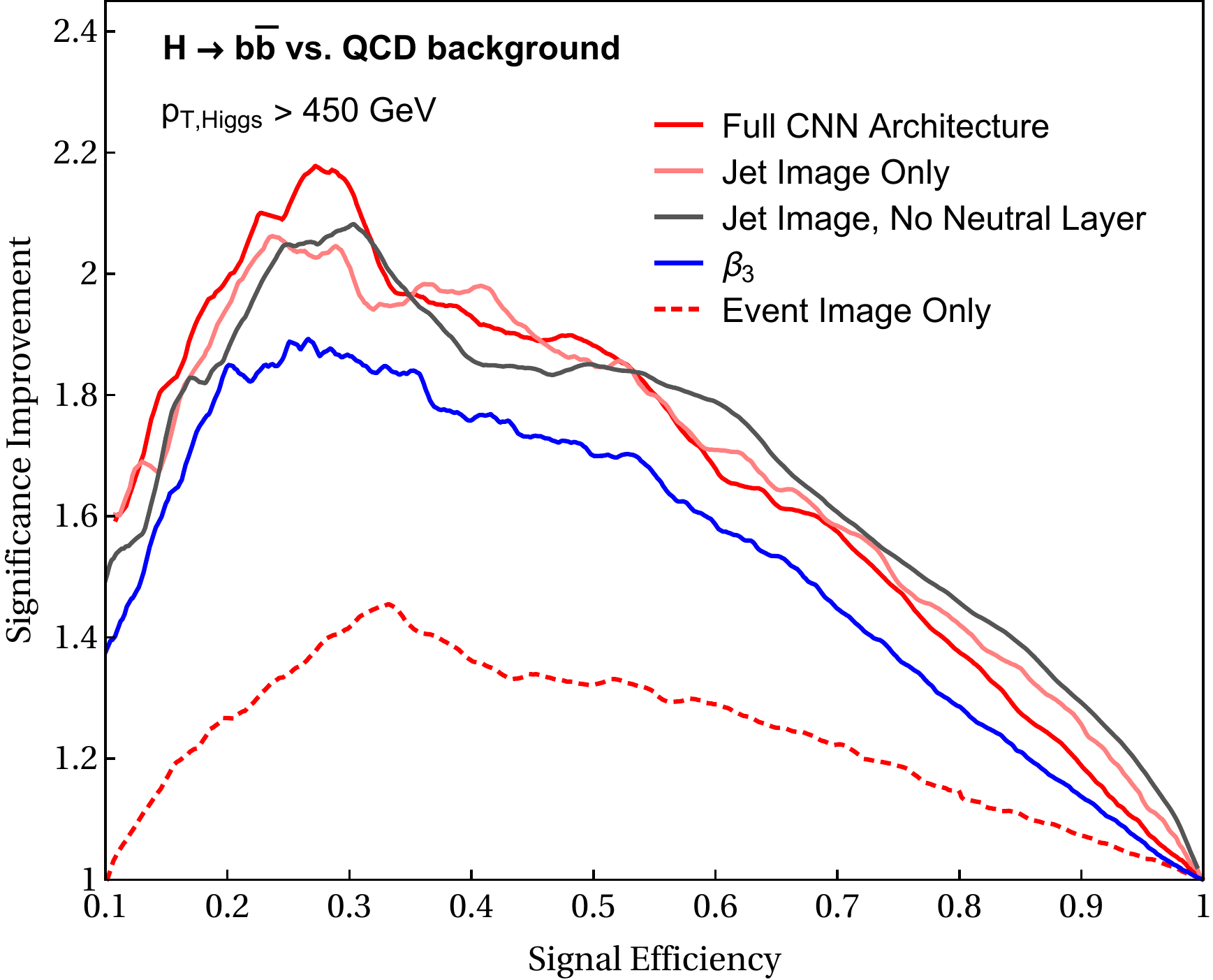}
  \end{center}
  \caption{Significance improvement characteristic (SIC) curve for various neural network setups as well as for the single observable $\beta_3$ proposed in \Ref{Datta:2017lxt}.}
  \label{fig:SIC_beta}
\end{figure}

The binned likelihood fit is performed in the mass range from \SIrange{50}{197}{\GeV} and using bins of width \SI{7}{\GeV}. (The CMS analysis performs the same fit in the  \SIrange{40}{201}{\GeV} with the same binning~\cite{Sirunyan:2017dgc}).  The corresponding SIC curve is shown in \cref{fig:SIC_beta}.  A maximum significance gain of about 2.2 is achieved with a signal efficiency of about \SI{25}{\percent}.  This means that if the significance with the nominal selection was 1 for a given dataset size, then after the application of the neural network, the new significance would be 2.2.  The maximum significance from the event stream only is about 1.4 while for the same value for the jet stream only is about 2.  This indicates that the jet information is much more important than the global information, though a significance gain of 1.4 is still important.  Since pileup is not included in the simulation, it is important to show that the performance is similar when pileup sensitive inputs are removed.  The ``no neutral layer'' curve in \cref{fig:SIC_beta} shows that the peak performance is robust and even better than the full network at high significance.  Intuitively, a network with more information should not be able to do worse, though in practice, this could occur due to weight sharing or from too few training examples.  For reference, the $\beta_3$ observable proposed in \Ref{Datta:2017lxt},

\begin{equation}
  \beta_3 = \frac{\pqty{\tau_1^{(0.5)}}^2 \pqty{\tau_2^{(1)}}^{0.5}}{\tau_2^{(2)}}\,,
\end{equation}
where $\tau_n^{(j)}$ is the $n$-subjettiness observable~\cite{Thaler:2010tr,Thaler:2011gf} with angular exponent $j$,  is also shown for comparison in \cref{fig:SIC_beta}. This single observable captures a significant fraction of the total significance improvement, but there is still more information available from the full two stream setup to boost the significance further.  A further investigation into the information learned by the network is described in \cref{sec:higgs_SM_ML_what_sub}.
 
\begin{figure}
  \begin{center}
    \includegraphics[width=8.5cm]{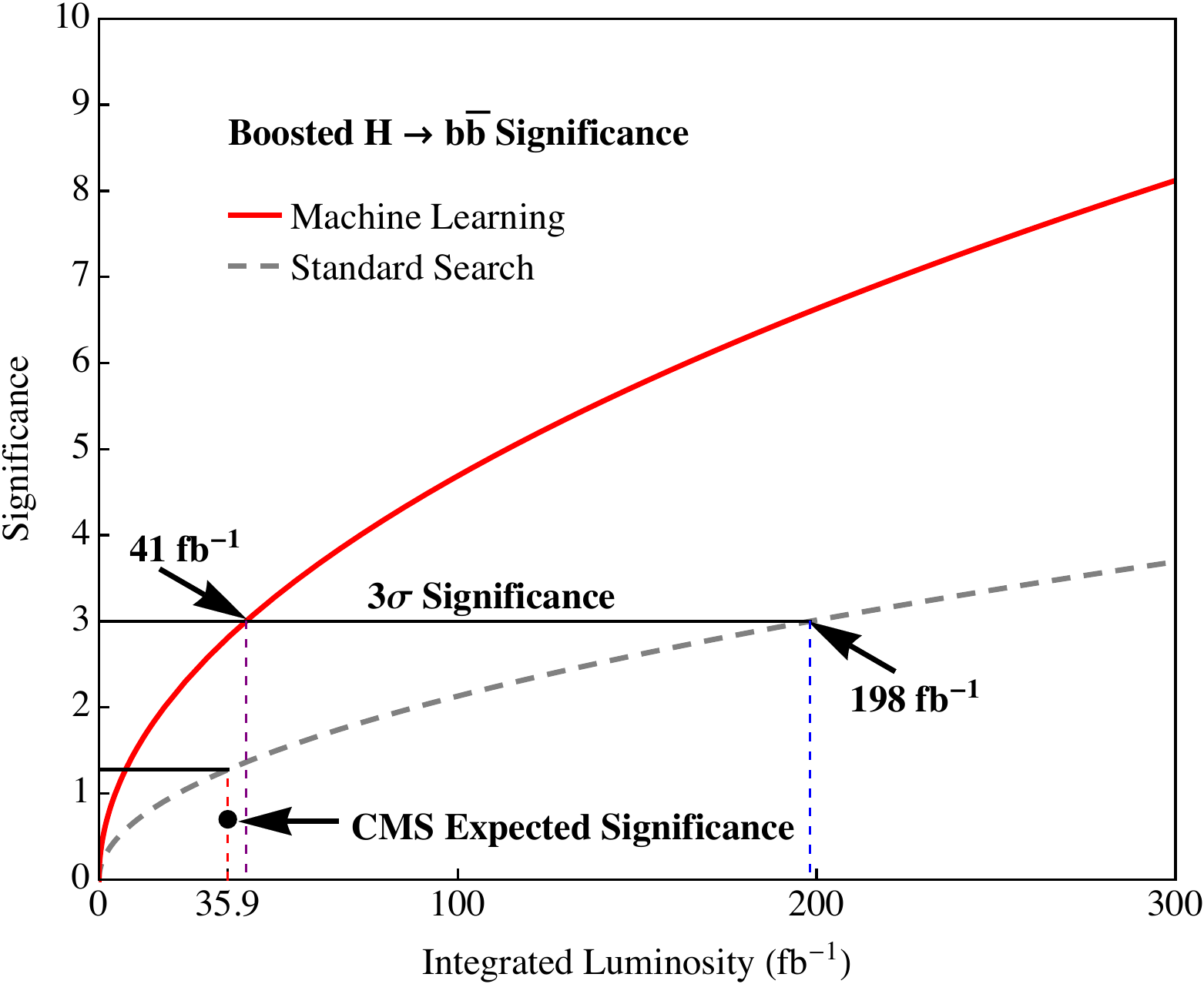}
  \end{center}
  \caption{The expected significance as a function of the integrated luminosity for the analysis with and without machine learning.  The vertical red dashed line corresponds to the dataset size from the current CMS result~\cite{Sirunyan:2017dgc} while the purple and blue dashed lines indicate the sizes required to reach $3\sigma$ with and without machine learning, respectively.  The full Run 2 (2015--2018) dataset will be about \SI{150}{\per\fb} and the full LHC dataset (up to 2023), prior to the HL-LHC, will be about \SI{300}{\per\fb}.}
  \label{fig:SIC}
\end{figure}

To understand the impact of a gain of 2.2 in the SIC, the expected significance for the SM $H \to b\bar{b}$ search is plotted as a function of integrated luminosity in \cref{fig:SIC}.  A center of mass energy of $\sqrt{s} = \SI{13}{\TeV}$ is assumed through the end of LHC Run 3, which corresponds to about \SI{300}{\per\fb}.  The curves follow the statical scaling of $\sqrt{\int L\, dt}$, where $L$ is the instantaneous luminosity.  The current CMS result reported an observed (expected) significance of 1.5 (0.7)~\cite{Sirunyan:2017dgc}.  As anticipated from the agreement with the mass distribution (\cref{fig:cms_verify}), the significance calculated using the simulation reported in \cref{sec:higgs_SM_standard} is very similar at 1.227. Without machine learning, ``evidence'' ($3\sigma$) may only be achieved after the full LHC dataset (up to 2023) and ``observation'' ($5\sigma$) may be possible only with the HL-LHC.  In contrast, with the application of the neural network, evidence may be achievable with the full Run 2 (2015--2018) dataset (about \SI{150}{\per\fb}) and observation may be possible well before the end of the LHC. This represents one of the main results of this paper, and emphasizes the possible gains to be had with ML.

\subsection{What is the Neural Network Learning?}
\label{sec:higgs_SM_ML_what}

With a significant improvement from the neural network, it is interesting to investigate in more detail what information the machine is exploiting beyond the existing search.  This section follows some of the procedures for such a study described in \Ref{deOliveira:2015xxd}.

\begin{figure}
  \captionsetup[subfigure]{labelformat=empty}
  \begin{center}
    \subfloat{\includegraphics[width=7.35cm]{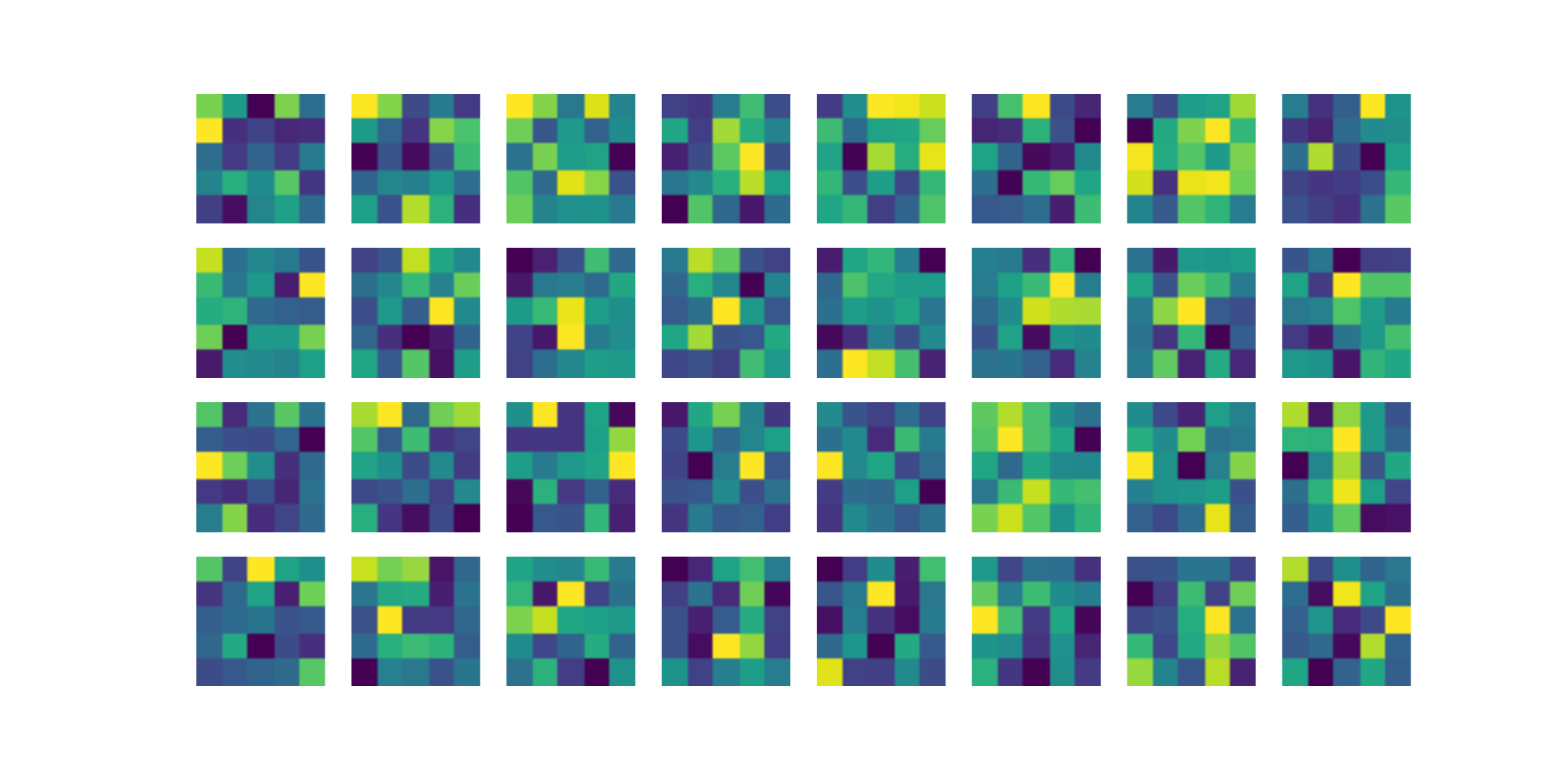}}
    \subfloat{\includegraphics[width=7.35cm]{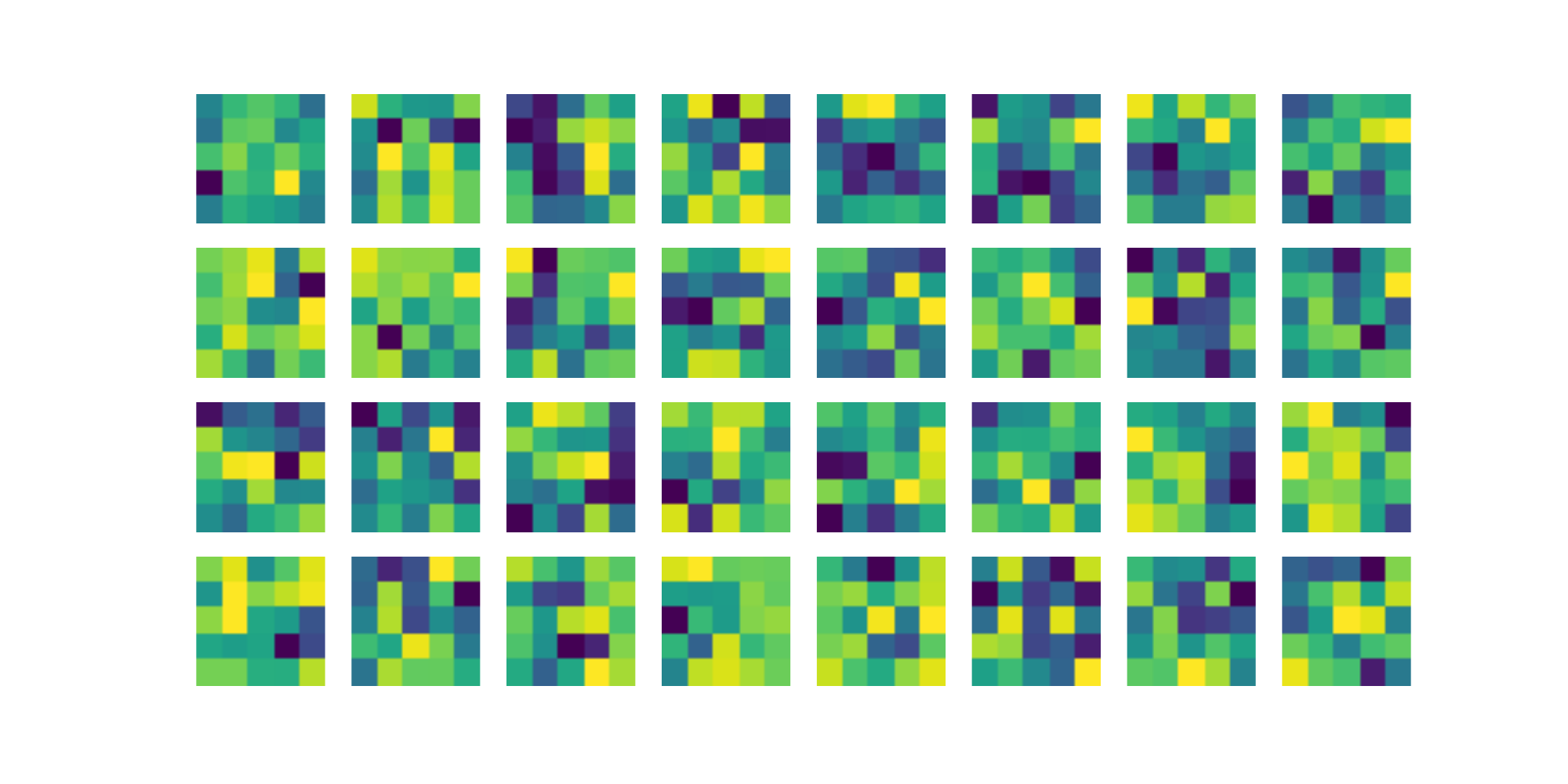}}
  \end{center}
  \begin{center}
    \subfloat{\includegraphics[width=7.35cm]{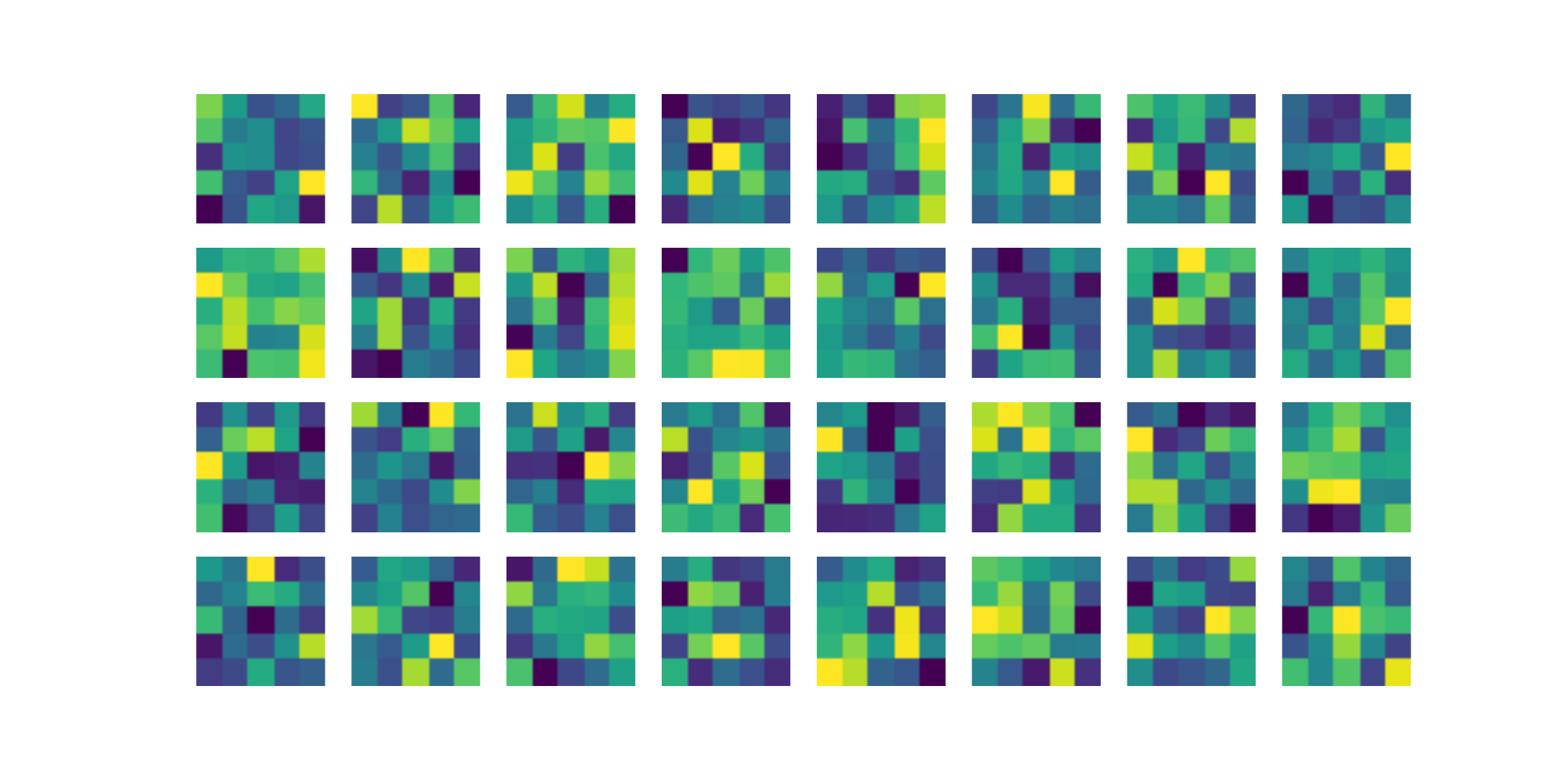}}
    \subfloat{\includegraphics[width=7.35cm]{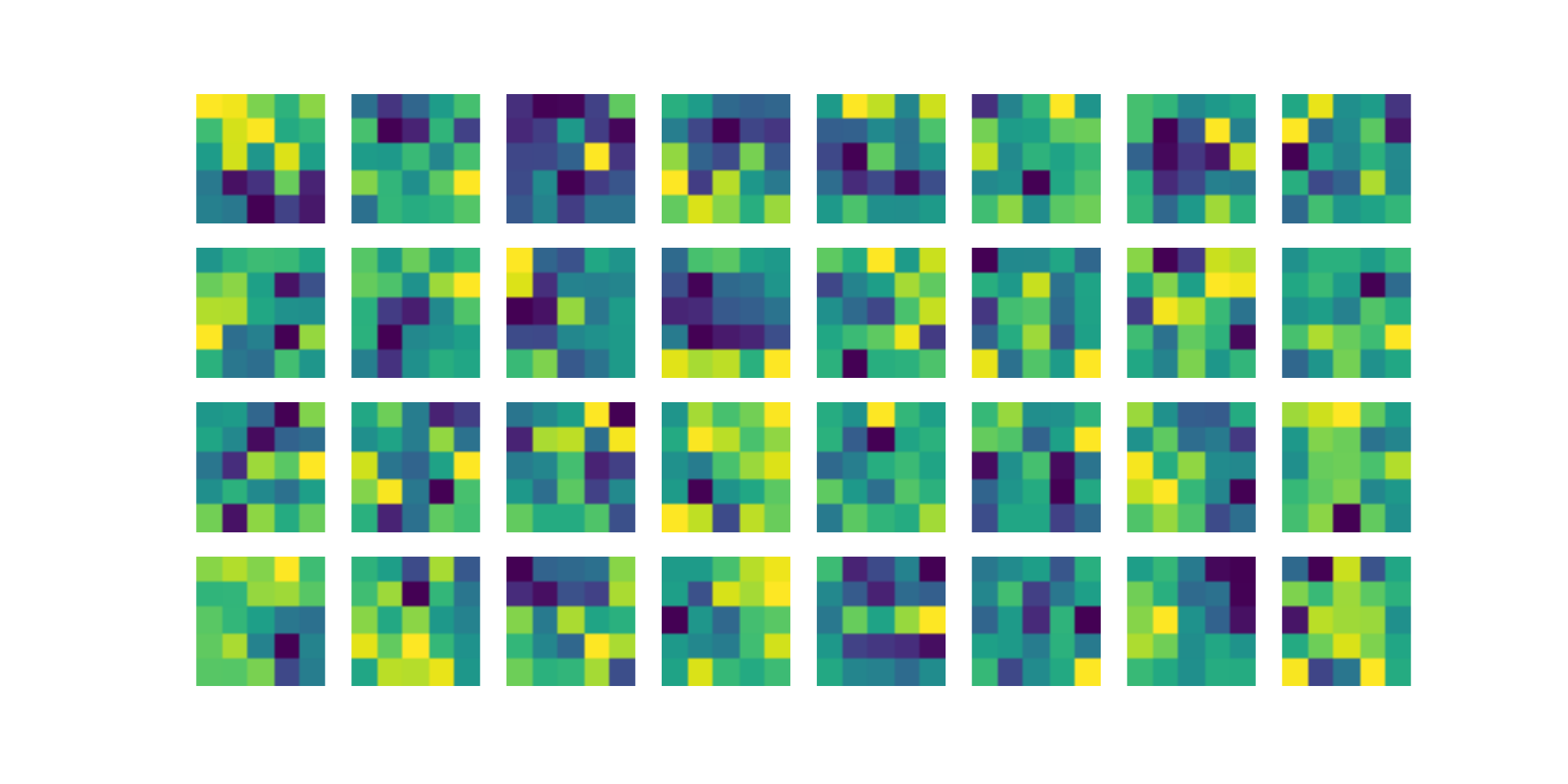}}
  \end{center}
  \begin{center}
    \subfloat[(a)]{\label{fig:filters_a}\includegraphics[width=7.35cm]{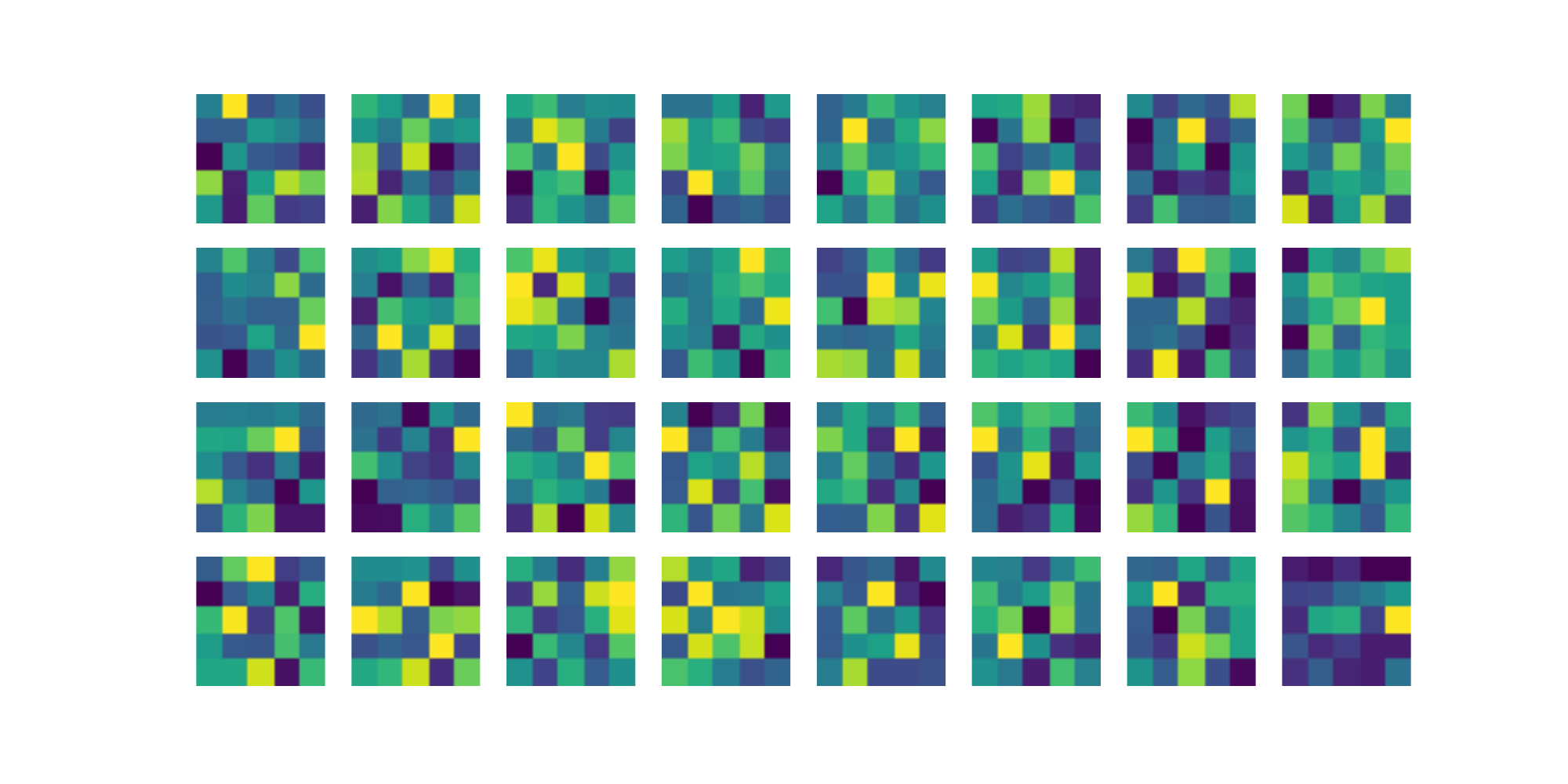}}
    \subfloat[(b)]{\label{fig:filters_b}\includegraphics[width=7.35cm]{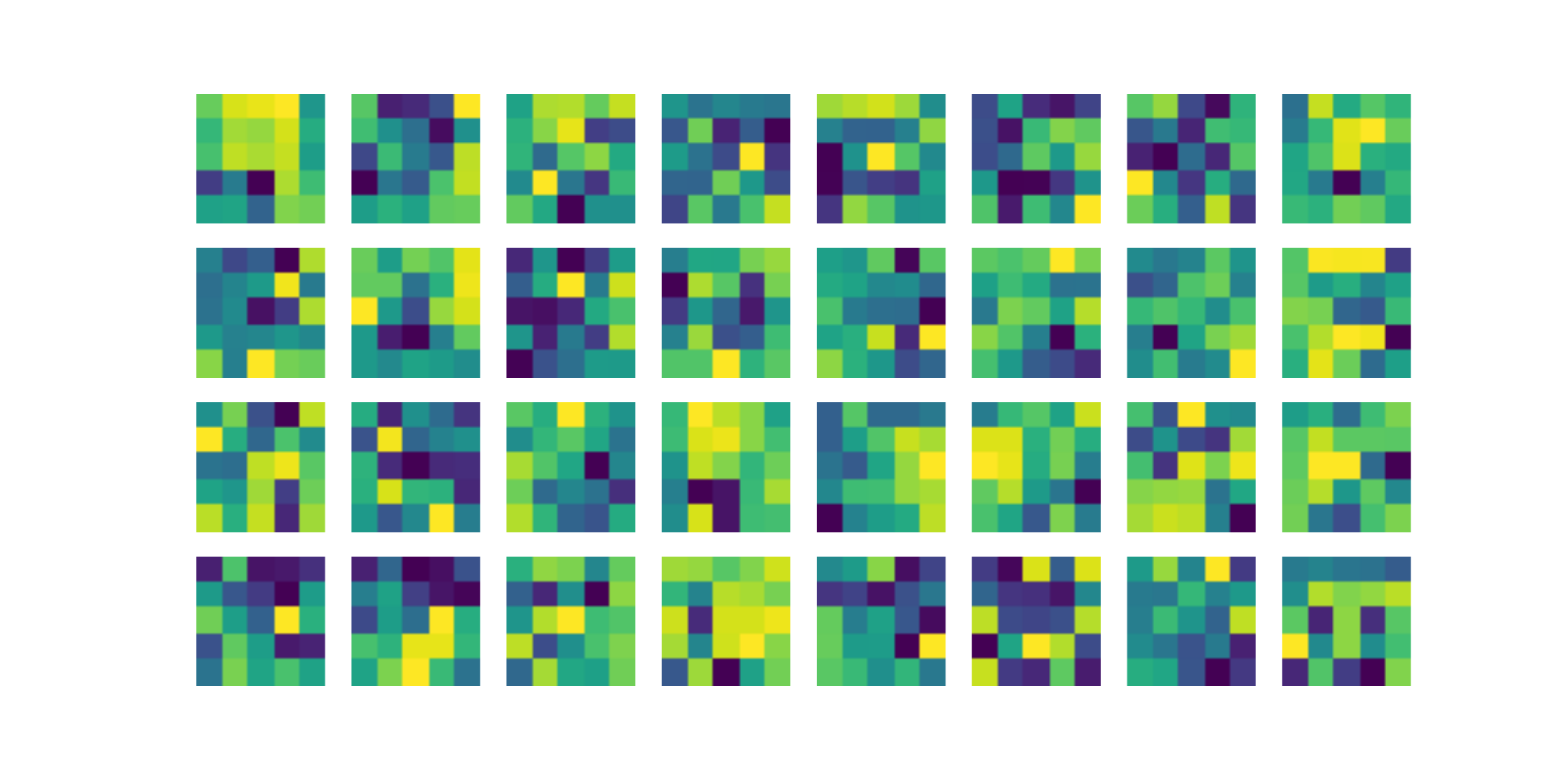}}
  \end{center}
  \caption{The 32 filters from the first layer of the total event CNN in (a) and the jet substructure CNN in (b).  The top row filters correspond to the charged $p_\text{T}$ layer, the second row shows the neutral $p_\text{T}$ layer and the bottom row is for the charged-particle multiplicity channel.}
  \label{fig:filters}
\end{figure}

First, \cref{fig:filters} shows the (first layer) convolutional filters from both streams of the network.  Since both streams are actually three-channel images, there are three sets of filters for each case.  While it is difficult to immediately recognize what the network is learning from these filters, there are some hints upon careful inspection.  In particular, the event images have a small number of ``hot spots.''  This may indicate that the network is learning to compute distances between prongs within jets and between jets.  In contrast, the jet image filters have many active pixels with complex shapes.  These filters are too small to span the typical subjet distance and so may be identifying the pattern of radiation between or around subjets.  The following sections examine the two streams of the network in more detail.

\subsubsection{Jet Substructure}
\label{sec:higgs_SM_ML_what_sub}

\begin{figure}
  \begin{center}
    \includegraphics[width=13.5cm]{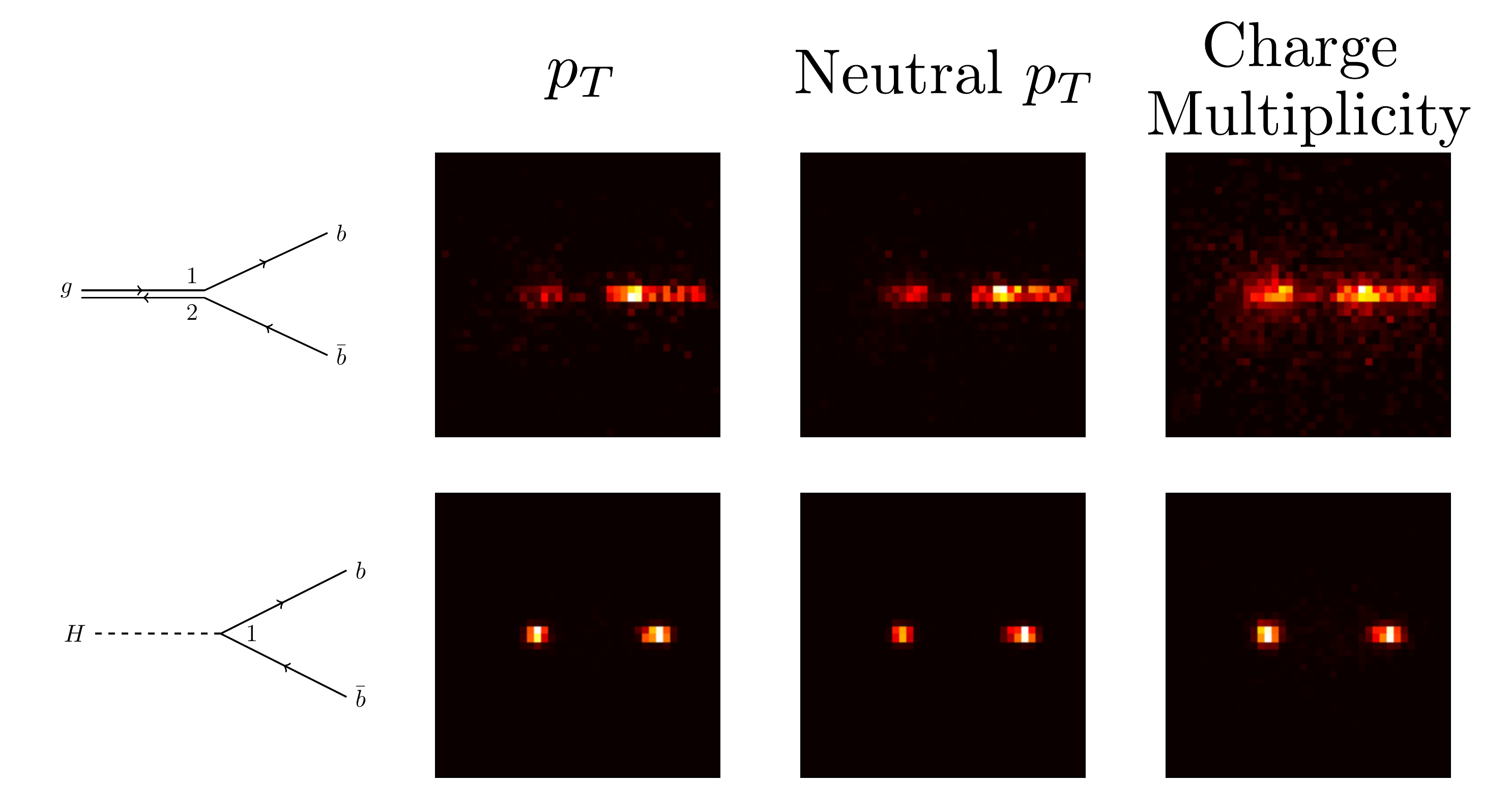}
  \end{center}
  \caption{Average jet images for the 100 most background like (top) and signal like (bottom) jets.  The jet images are weighted by the \pt in the first column, the neutral \pt in the second column, and the charge multiplicity in the third column. Due to the different color flows, the signal like ($H \to b\bar{b}$) jets have a more contained color flow pattern.}
  \label{fig:jet_images}
\end{figure}

\begin{figure}
  \begin{center}
    \includegraphics[width=6.5cm]{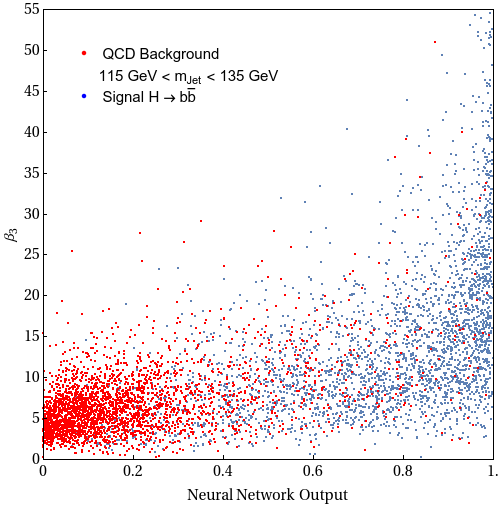} \\\vspace{3mm} \includegraphics[width=13.5cm]{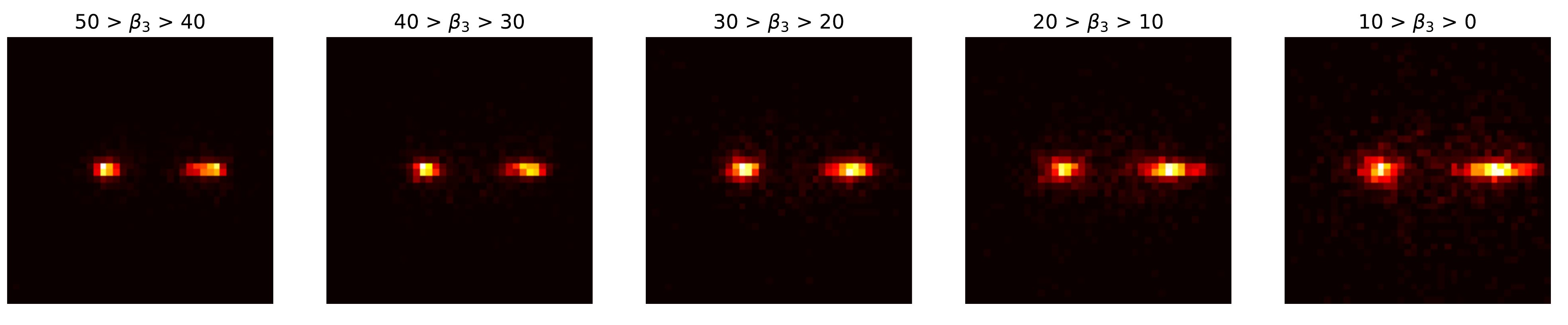}
  \end{center}
  \caption{Top: the joint distribution of $\beta_3$ and the neural network.  Bottom: the average of many signal images in decreasing $\beta_3$ intervals.}
  \label{fig:jet_images2}
\end{figure}

\begin{figure}
  \begin{center}
    \includegraphics[width=13.5cm]{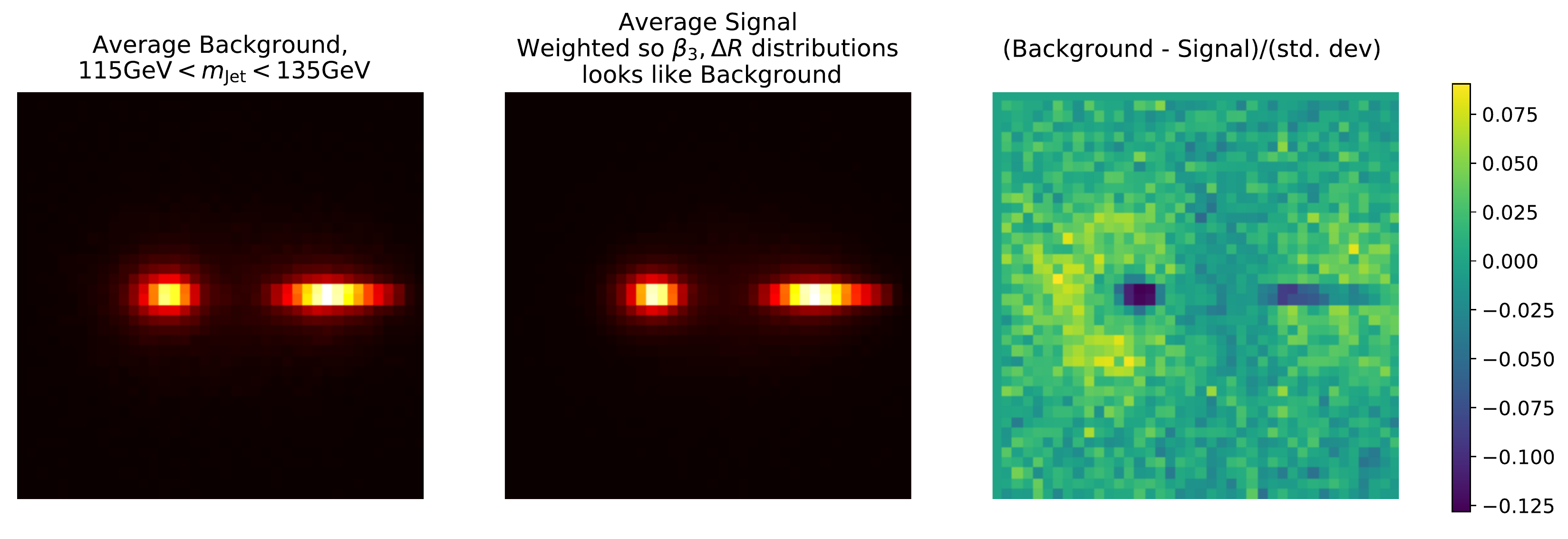}
  \end{center}
  \caption{Left (middle): The average background (signal) images after reweighting the $\beta_3$ distribution and the $\Delta R$ (between b-tagged subjets) distribution in signal to be the same as in background.  Right: The average difference between the left and middle images.}
  \label{fig:jet_images3}
\end{figure}

As emphasized earlier, the $H \to b\bar{b}$ search is different from other boosted hadronically decaying massive boson studies because the application of double $b$-tagging already enforces a two-prong topology.  Therefore, two-prong tagging is not as useful.  Studies to further optimize the event selection with $N_2$ confirm this expectation --- little significance gain is possible using only this state-of-the-art two-prong tagging technique (see also \Ref{ATL-PHYS-PUB-2015-035}).  One of the attractive features of jet images is that they can be directly inspected to visualize the information content.  For example, \cref{fig:jet_images} shows the average of the 100 most signal-like and most background-like jets, according to the neural network.  The two-prong structure of both signal and background is clear in all three channels.  The main difference between $gg\rightarrow b\bar{b}$ and $H\to b\bar{b}$ is the orientation of the radiation between and around the two prongs.  As expected due to the different color structure, the radiation pattern around the two prongs is more spread out for the gluon case.  \Cref{fig:jet_images2} shows additional images that are split by their value of $\beta_3$.  It is clear from the images that low $\beta_3$ values (background-like) pick out subjets with a broader radiation patterns compared with high $\beta_3$ (signal-like) images.  However, the top plot of \cref{fig:jet_images2} clearly indicates that $\beta_3$ is not the same as the neural network, so there is additional information to learn.  \Cref{fig:jet_images3} tries to visualize the additional information.  The distribution of $\beta_3$ in the signal is reweighted to be the same as the background so that $\beta_3$ by itself is not useful for discrimination.  The average images for signal and background look very similar by eye, but the difference of the average images reveals interesting structure.  These structures still show an enhanced radiation pattern around the subjets for the background relative to the signal --- there is thus more color flow information available to learn than is captured by $\beta_3$ alone.

Drilling down into the information content of the jet images in more detail, perhaps using more of the techniques from \Ref{deOliveira:2015xxd} and understanding to what extent $\beta_3$ captures color flow and other effects is of great interest for future studies.\footnote{Explicit color flow tagging observables have been proposed in the past, such as the jet pull~\cite{Gallicchio:2010sw}.  However, these observables are not powerful tools for tagging~\cite{ATLAS-CONF-2014-048,deOliveira:2015xxd}, though they are very useful for precision studies~\cite{Aad:2015lxa,Aaboud:2018ibj}.  We have checked and the SIC for jet pull does not significantly exceed unity.}

\subsubsection{Global Event}
\label{sec:higgs_SM_ML_what_global}

While much attention has been devoted to the extraction of information from jet substructure, less has been payed to the extraction of discrimination power from the full event. At the same time, probing what is learned from event properties is more complicated than for the jet image due to the reduced symmetry.  As with the jet substructure, due to the color singlet nature of the Higgs, we expect that the color flows in signal and background jets should be distinct. From our study, we find that while this information does not provide as much discrimination power as the jet substructure, it nonetheless provides an additional gain in significance. While several observables for discriminating global color flow have been proposed~\cite{Gallicchio:2010sw,Krohn:2011zp,Ebert:2016idf,Harland-Lang:2016vzm}, this is in general quite a challenging task. Furthermore, we expect that it would be quite topology dependent. Nevertheless it would be interesting study in more detail, since it has not received much attention. We believe that ML is an ideal technique for extracting complicated global event information that has not yet been exploited to its full potential in LHC searches.

\begin{figure}
  \begin{center}
    \includegraphics[width=7cm]{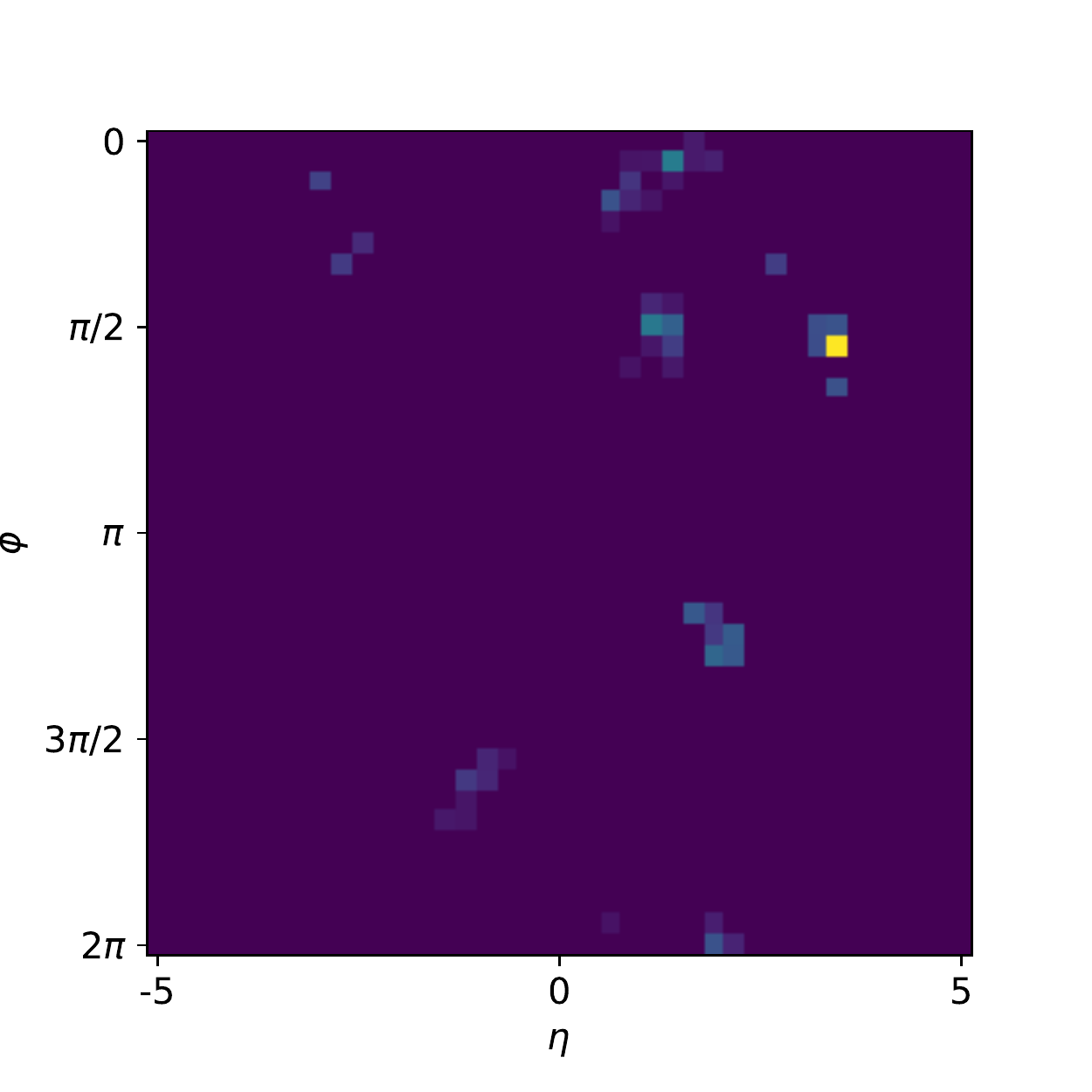}\hspace{5mm}\includegraphics[width=6.5cm]{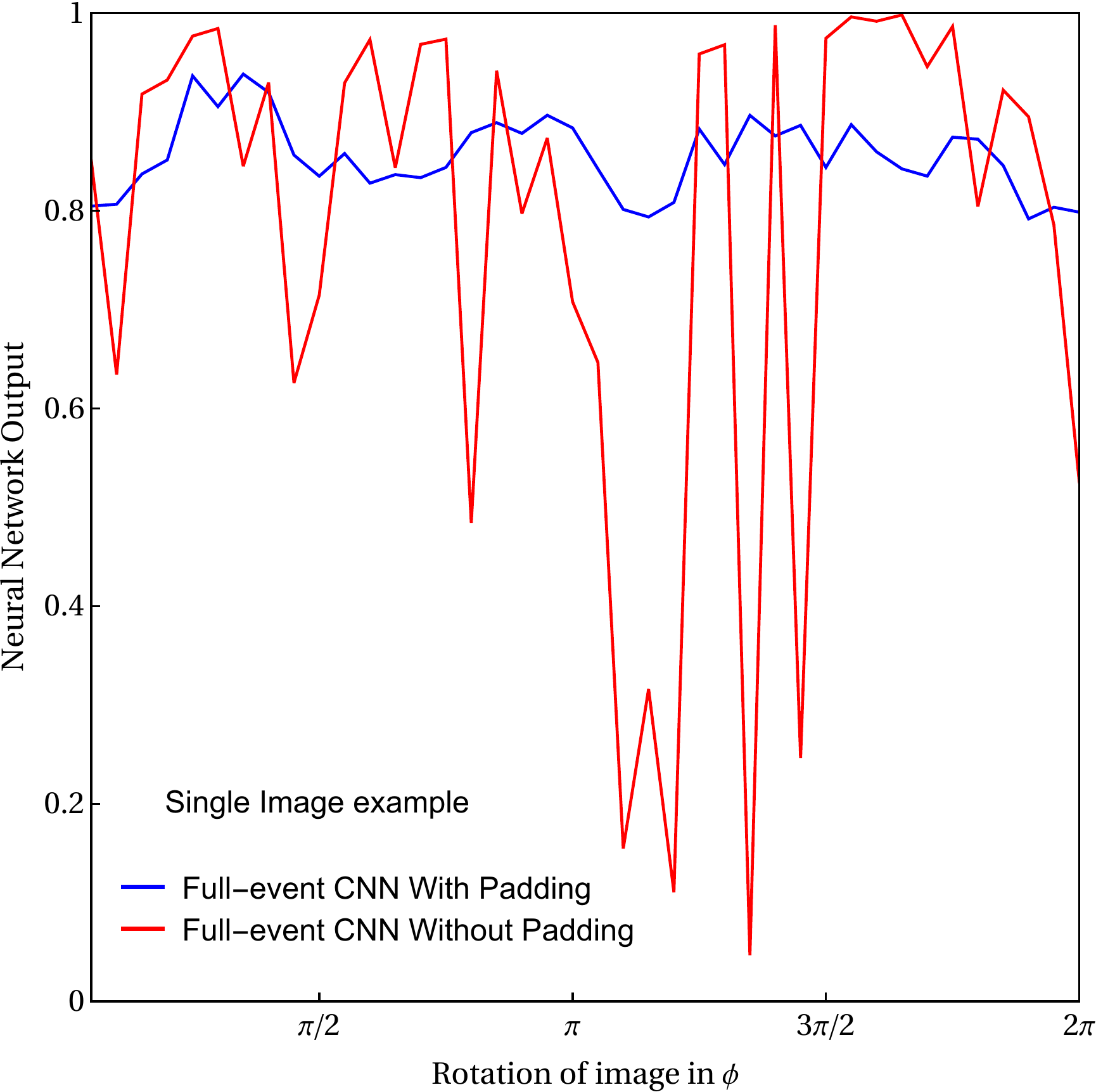}
  \end{center}
  \caption{Left: A typical signal event image.  Right: The output of the neural network on the left image, after rotation in the $\phi$ direction by the given number of pixels.}
  \label{fig:jet_images4}
\end{figure}

\begin{figure}
  \begin{center}
  \includegraphics[width=7.5cm]{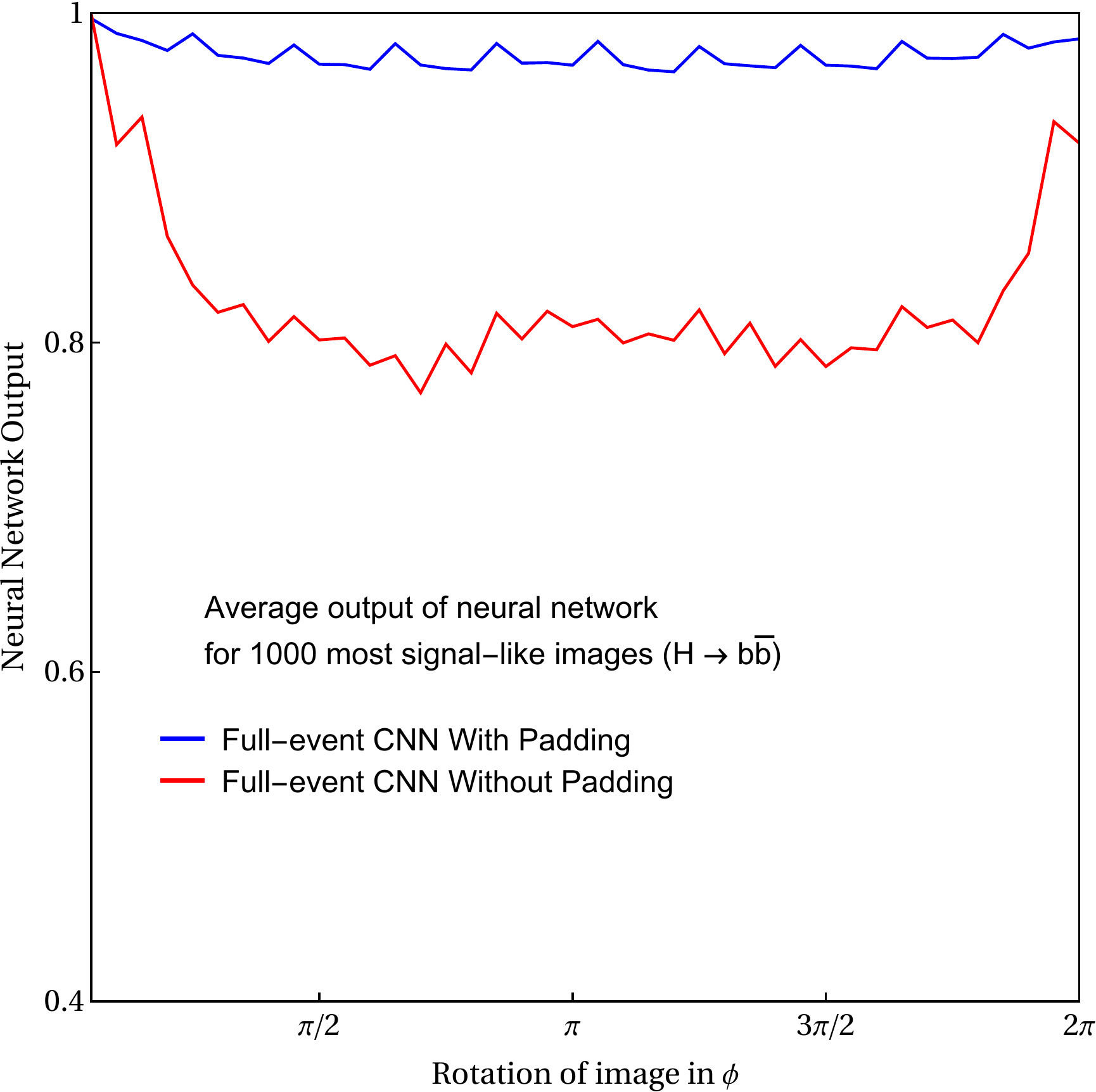}
  \end{center}
  \caption{The average signal neural network output after rotating the images by a fixed amount.  }
  \label{fig:jet_images5}
\end{figure}

We also highlight the efficacy of the padding that renders the convolutional layers of the neural network symmetric under rotations in $\phi$ by one pixel.  This is a new feature of our neural network, which we find to be helpful for training stability.  \Cref{fig:jet_images4} shows a typical signal image and how the neural network output changes as the event is rotated in $\phi$.  As desired, the network with the padding at every convolution layer is much more stable than the ones without the padding.  The reason that the padded network is not completely invariant under rotations in $\phi$ is that the dense layers at the end of the network break the $\phi$ symmetry while for rotations as scales below a single pixel the discretization breaks the invariance.  \Cref{fig:jet_images5} shows a similar trend after averaging over many events.  

\subsection{Recommendations for Future Searches}
\label{sec:higgs_SM_reco}

Due to the important of the $H \to b\bar{b}$ channel for probing the Higgs sector at the LHC, we conclude this section with some concrete recommendations for improving the LHC searches, reiterating the points found in this section. In particular, although the most power is gained from a neural network, we have shown that a large fraction of this information can be obtained through simple observables, which can immediately be implemented in current searches for boosted $H \to b\bar{b}$.  A neural network using charged information could also be applied without requiring extensive calibration studies.  A key component of what the network can learn since the signal and background are already in a two-prong topology is the color flow.  Quantifying the additional information in the form of compact analytic observables is an interesting and important part of future work.

\section{\boldmath High-\texorpdfstring{\pt}{pT} Higgs for BSM Physics}
\label{sec:BSM}

Beyond the discovery of the $H \to b\bar{b}$ decay, a major motivation for the study of boosted $H \to b \bar{b}$ final states in particular is that it allows one to study the structure of the $gg \to H$ process at high \pt. While in the Standard Model this is primarily due to the contribution of a virtual top quark loop, the total cross section $\sigma(gg \to H)$ is only sensitive to the low-energy limit of this loop, in which it is extremely well approximated by a dimension-five operator with no dependence on $m_t$. At $\pt \gtrsim 2m_t$, this is no longer true, as the physical momentum running through the loop is comparable to $m_t$, allowing potential new physics contributions to the loop to be disentangled that are not observable for the total cross section by observing the \pt dependence. This general observation has been explored in \Refs{Grojean:2013nya,Azatov:2013xha,Buschmann:2014twa,Schlaffer:2014osa,Azatov:2016xik}. In this section we apply our machine learning techniques and illustrate how the improved significance for $H \to b\bar{b}$ translates to improved bounds on BSM physics.

We are interested in probing new physics in the $gg \to H$ production loop that can be modeled as dimension-6 operators. Following \Ref{Grojean:2013nya}, the operators modifying $gg \to H$ production cross section are parameterized as 
\begin{multline}
  \mathcal{L}_\text{eff} = \mathcal{L}_\text{SM}
                           + \pqty{c_y \frac{y_t}{v^2} |H|^2 \bar{Q}_L \tilde{H} t_R + \text{h.c.}}
                           + c_H \frac{1}{2v^2} \partial_\mu |H|^2 \partial^\mu |H|^2 \\
                           + c_g \frac{\alpha_s}{12\pi v^2} |H|^2 G^a_{\mu\nu} G^{a\mu\nu}
                           + \tilde{c}_g \frac{\alpha_s}{8\pi v^2} |H|^2 G^a_{\mu\nu} \widetilde{G}^{a\mu\nu}.
\end{multline}
Here $G^{a\mu\nu}$ is the QCD field strength, and $\widetilde{G}^{a\mu\nu} = \frac{1}{2} \epsilon^{\mu\nu\sigma\rho} G^a_{\sigma\rho}$ its dual. After electroweak breaking, the induced operators affecting the coupling of the Higgs boson to tops and gluons take the form
\begin{equation}
  \mathcal{L}_\text{eff} = \mathcal{L}_\text{SM}
                           - \kappa_t \frac{m_t}{v} t\bar t h
                           + i\tilde{\kappa}_t \frac{m_t}{v} \bar t \gamma_5 t h
                           + \kappa_g \frac{\alpha_s}{12\pi}\frac{h}{v} G^a_{\mu \nu} G^{\mu \nu a}
                           + \tilde{\kappa}_g \frac{\alpha_s}{8\pi}\frac{h}{v} G^a_{\mu \nu} \tilde{G}^{a \mu \nu}\,,
\end{equation}
where 
\begin{equation}
  \kappa_t = 1 - \text{Re}(c_y) - \frac{c_H}{2} \qcomma
  \kappa_g = c_g \qcomma
  \tilde{\kappa}_t = \Im(c_y) \qcomma
  \tilde{\kappa}_g =\tilde{c}_g\,,
\end{equation}
so that one degeneracy between $c_H$ and the real part of $c_y$ remains. In the following we will only be interested in $CP$-even terms, and explicitly set $\tilde{\kappa}_t = \tilde{\kappa}_g = 0$, implicitly demanding $c_y$ be real.

\begin{figure}
  \begin{center}
    \includegraphics[width=8.5cm]{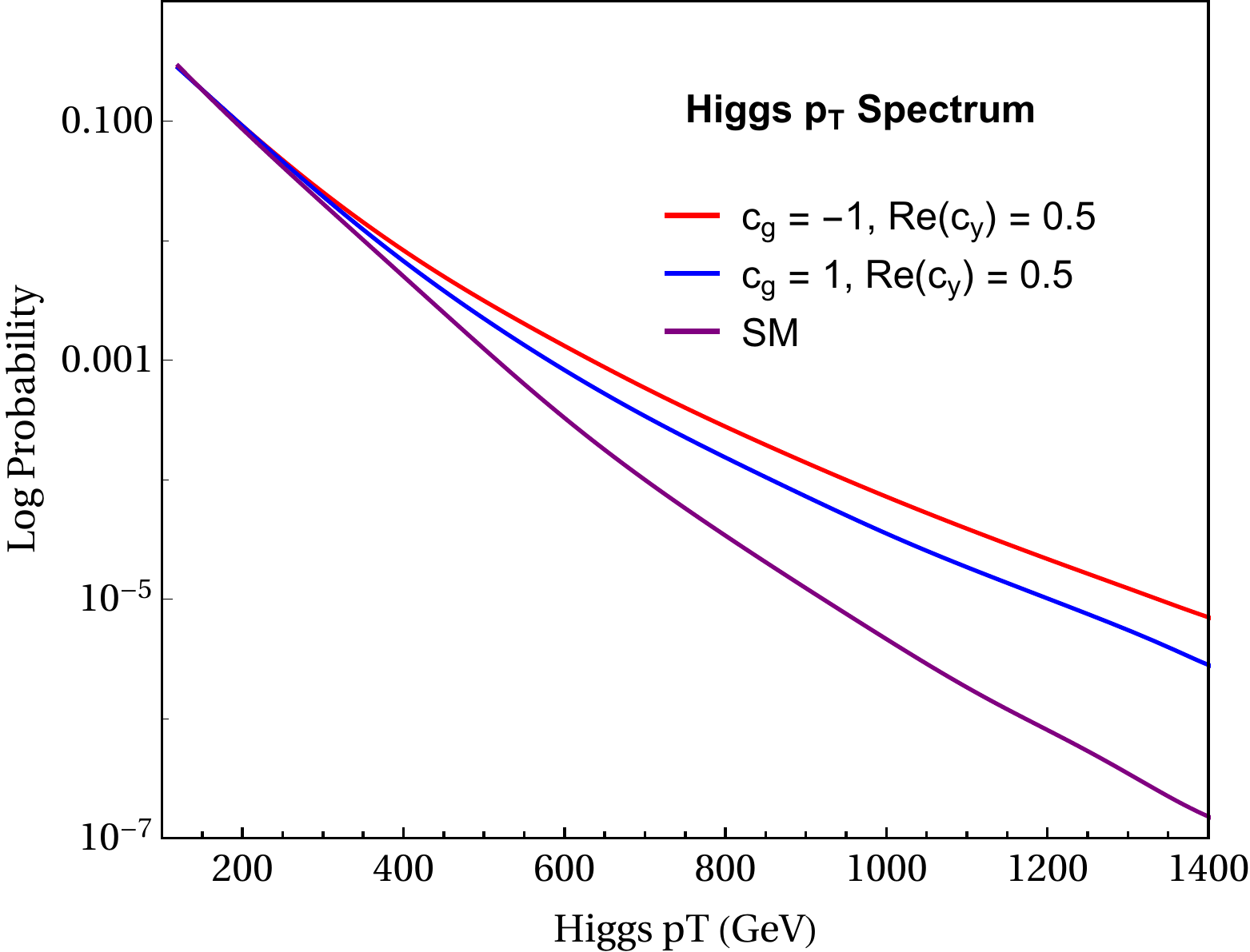}
  \end{center}
  \caption{Typical hardening of the \pt spectrum of the Higgs for nonzero values of the Wilson coefficient $c_g$, defined in the text.}
  \label{fig:hard}
\end{figure}

Although the two couplings that we are interested in probing have distinct physical effects, namely $\kappa_t$ acts as a correction to the top Yukawa, while $\kappa_g$ corrects the $ggH$ coupling, the Higgs low energy theorem \cite{Ellis:1975ap,Shifman:1979eb} guarantees that they contribute to the inclusive Higgs production cross section as $(\kappa_t+\kappa_g)^2$ up to corrections of $\ord(m_h^2/m_t^2)$. As shown in \Ref{Grojean:2013nya}, this degeneracy is broken by the cross-section for $H + \text{jet}$ production for a given $\pt^\text{min}$ cut on the Higgs scales like
\begin{equation}
  \frac{\sigma_{\pt^\text{min}}(\kappa_t,\kappa_g)}{\sigma_{\pt^\text{min}}^\text{SM}}
    = (\kappa_t + \kappa_g)^2 + \delta \kappa_t \kappa_g + \epsilon \kappa_g^2\,,
\end{equation}
where $\epsilon$ and $\delta$ are terms dependent on the $\pt^\text{min}$ cut placed on the Higgs. These are given in \Ref{Grojean:2013nya} for tabulated values of $\pt^\text{min}$, computed at one-loop order with full $m_t$ dependence. For $\pt^\text{min} \lesssim 2m_t$, behavior similar to the inclusive rate is observed as $\epsilon, \delta \ll 1$, while for $\pt^\text{min} \gtrsim 2m_t$, $\epsilon, \delta$ become of $\ord(1)$, leading to a hardening in the \pt spectrum of the BSM model as compared to the SM model after taking into account the overall scaling of the cross-section given by the $(\kappa_t + \kappa_g)^2$ term (provided that $c_g \neq 0$). We can see this effect in \cref{fig:hard}. Consequently, performing a search for boosted Higgs can provide bounds on the Wilson coefficients $\kappa_g  = c_g$ and the combination $\kappa_t = 1 - \Re(c_y) - c_H/2$. 

\subsection{Simulation}
\label{sec:sim_bsm}

Our signal of interest is dominated by the interference of the SM $gg \to Hj(j)$ process, with the higher dimensional operators given above. Since \mg{} is currently unable to compute the interference effects of processes that start at loop-level, a modification of the typical \mg{} procedure is necessary to correctly generate the processes we want to study. The effect of the operators parameterized by $c_g$ and $\tilde{c}_g$ are recovered by implementing a fictitious heavy top partner whose mass is set to be large enough that a contact operator approximation remains valid for all LHC processes (nominally \SI{10}{\TeV}) and whose coupling to the Higgs is tuned to give the correct dependence on the high-dimension operator coefficient. The other operators are implemented as actual higher-dimension operators, as is conventional.

\begin{figure}
  \begin{center}
    \includegraphics[scale=0.4]{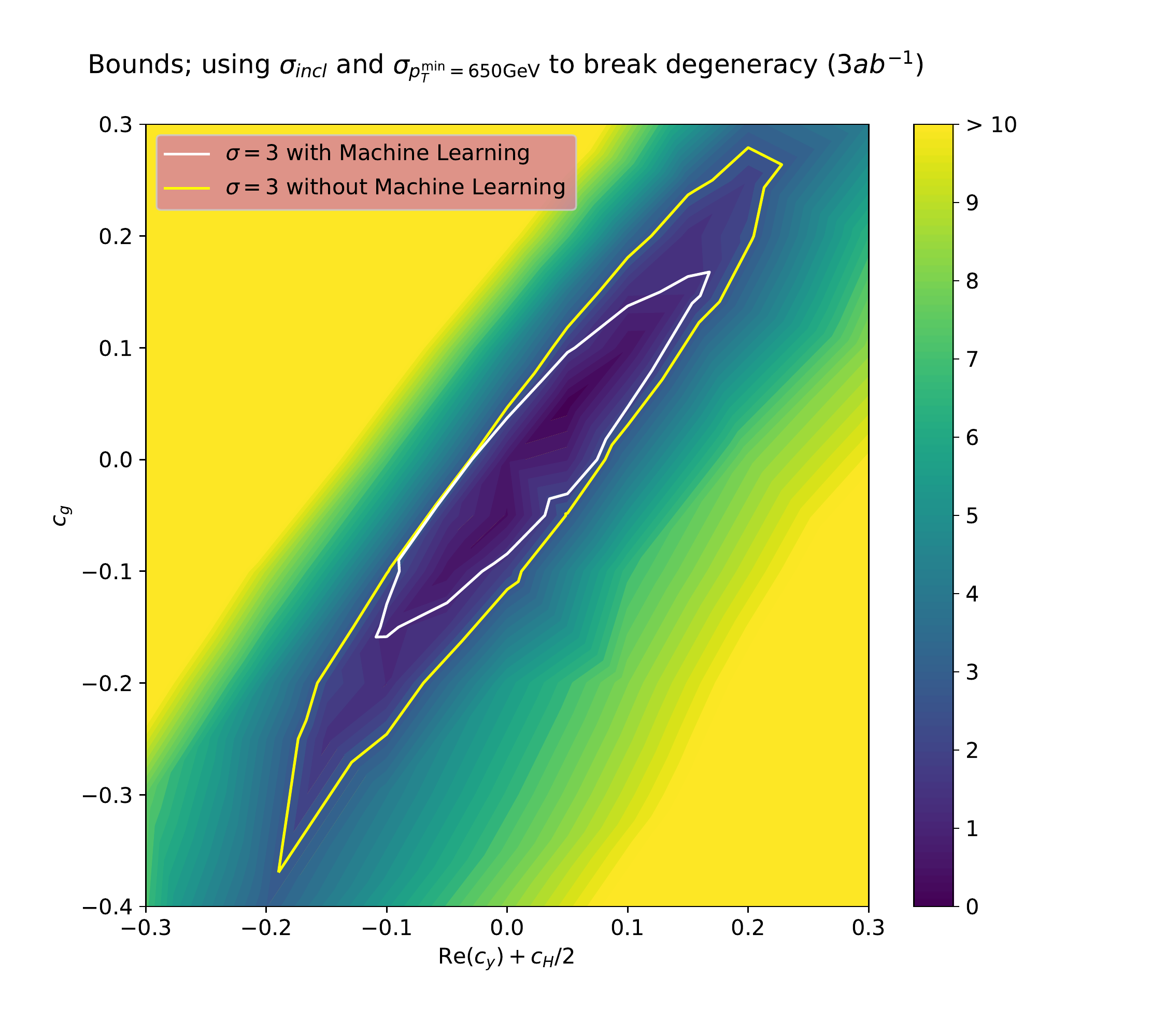}
  \end{center}
  \caption{Expected constraints on the couplings $c_g$, $c_y$ and $c_H$ derived with and without ML. A significant improvement along the anticorrelated direction of $c_g$ and $\Re(c_y) + c_H/2$ is observed using ML.}
  \label{fig:BSMfinal}
\end{figure}

\subsection{Results and interpretation}
\label{sec:res_bsm}

We apply our two-stage convolutional neural network to improve the bounds on the couplings $c_g$ and the combination $\Re(c_y) + c_H/2$ as compared with the standard search. In \cref{fig:BSMfinal} we show the constraints from comparing the inclusive cross section to one with $\pt^\text{min} = \SI{650}{\GeV}$. We see significant gains using machine learning, corresponding to the improved significance for the Higgs seen earlier.

To properly situate these results, we summarize the current bounds and future prospects for constraining these operator coefficients in the absence of a dedicated high-\pt Higgs analysis. While theoretical constraints from general principles such as causality and locality do exist~\cite{Low:2009di} (in particular $\Re(c_y) + c_H/2 > 0$ seems to always be true when generated within quantum field theory), the best current bounds on the couplings $c_g$ and $\Re(c_y) + c_H/2$ come from a combination of the most recent inclusive Higgs cross section measurements~\cite{Aaboud:2017vzb} and recent global fit of the Standard Model Effective Field Theory to all current Higgs and electroweak data performed in \Ref{Ellis:2018gqa}. The inclusive Higgs measurement constrains a combination of couplings for which the linearization $c_g - \Re(c_y) - c_H/2$ is an excellent approximation to be \num{0.29(46)} at $3\sigma$ using \SI{36.1}{\per\fb} of data.  The global fit provides current world averages (again with $3\sigma$ uncertainties) of $c_g = \num{0.10(30)}$,  $\Re(c_y) = \num{-4.7(78)}$, $c_H = \num{-1.1(18)}$, all consistent with zero at $2\sigma$. These results are primarily driven by the LHC Run II Higgs measurements, all using \SI{36.1}{\per\fb} (\SI{35.9}{\per\fb}) of data from ATLAS (CMS), although since the effect of possible other higher-dimension operators on the backgrounds is not included these bounds should be interpreted with care. The dominant discriminating power is provided by looking for deviations in the $h \to WW^*, ZZ^*$ decays, with the most constraining bounds clearly being on the coefficient $c_g$.\footnote{A recent combined fit from CMS using high-\pt $H \to b\bar{b}$ and well as differential $H \to \gamma\gamma$, and $H \to ZZ^* \to 4\ell$ decays~\cite{CMS-PAS-HIG-17-028} finds bounds of $c_g \simeq 0.12 \pm 0.42$, $|\Re(c_y) + c_H/2| \lesssim 0.5$ using our conventions. These are nearly competitive with the global fits on $c_g$ and would clearly improve the fits of the other bounds if included in the global average. However, they also assume the Higgs branching ratios maintain their Standard Model values and are significantly weaker if these ratios are allowed to float.}

Conservatively assuming no improvement in the treatment of systematic or theoretical errors, analogous bounds to those discussed above with a full \SI{3}{\per\atto\barn} dataset should be able to reduce uncertainties by a factor of 2 in the inclusive Higgs cross section linear combination and by \SIrange{20}{25}{\percent} for the global fits. Comparing this to the projections of \cref{fig:BSMfinal}, our proposed analysis has the potential to exceed these conservative extrapolations on sensitivity by a factor of a few.

\section{Conclusions}
\label{sec:conclusions}

In this paper, we have applied modern machine learning techniques to improve the search for the $H \to b\bar{b}$ decay at the LHC. This decay offers a powerful probe of BSM contributions to the $gg \to H$ loop at high \pt.  Using our techniques, this process may be discoverable at the LHC (prior to the HL-LHC).  

A new feature of our analysis is that we have used a two stream convolutional neural network, with one stream acting on the double $b$-tagged jet, and the other stream acting on the global event information. This enables us to not only exploit the maximal information in the event, combining both jet substructure information and global information, but also allows us to more easily identify the dominant physics features that the neural network is learning. In particular, we find that a significant fraction of this information is not contained in the recently proposed $\beta_3$ observable. Disentangling these differing sources of information is challenging in standard analyses, which substructure observables nominally designed to identify two-prong substructure, although in the course of optimization they may become sensitive to other features as well.  Resolving an event at multiple scales and in various regions of phase space is a generic technique that should enable significant improvements in other LHC searches.  By probing the neural network in detail, it may also be possible to use the neural networks as a guide to building compact, analytical, simple observables that nearly saturate the machine learning performance.  With such tools in hand, increasingly extreme regions of phase space can be thoroughly explored. 

\acknowledgments

We thank the LBNL theory and experimental groups, especially Heather Gray, for lively discussions on this topic at the weekly ``cookie time.'' 
We thank Kaustuv Datta, Phil Harris, Andrew Larkoski, and Caterina Vernieri for comments and suggestions on the analysis and manuscript.  We also thank Wahid Bhimji and Steve Farrell at the National Energy Research Scientific Computing Center (NERSC) for helpful discussions, Phil Harris for useful discussions and technical help on the sample generation, Olivier Mattelaer for help with \mg{}, and Frank Tackmann for stimulating discussions at Les Houches 2017 that started this project.  Finally, we are grateful for the opportunity to use the CORI and EDISON supercomputing resources at NERSC.  This work was supported by the U.S.~Department of Energy, Office of Science under contract DE-AC02-05CH11231. MF is supported by the U.S.~Department of Energy under contract DE-SC0011640.  JL is supported by the Center for Computational Excellence, a project funded by the Computational HEP program in U.S.~Department of Energy Office of Science.

\bibliographystyle{jhep}
\bibliography{myrefs}
\end{document}